\documentclass[12pt]{article}
\usepackage[utf8]{inputenc}
\usepackage[left=1in, right=1in, top=1in]{geometry}
\usepackage[utf8]{inputenc}
\usepackage{appendix}
\usepackage{mathtools}
\usepackage{amsmath}
\usepackage{amssymb}
\usepackage{enumitem}
\usepackage{comment}
\usepackage{amsthm}
\usepackage{setspace}
\usepackage{graphicx}
\usepackage{color}
\usepackage{braket}
\usepackage{bbold}
\usepackage{url}
\usepackage{bibentry}
\usepackage{amsmath}
\usepackage{amsfonts}
\usepackage{clipboard}
\usepackage{cite}
\usepackage[colorlinks=true, linkcolor = blue, citecolor = blue, urlcolor = blue,linktocpage=true]{hyperref}
\usepackage{amssymb}
\usepackage{graphicx}
\usepackage{float}
\usepackage{caption}
\numberwithin{equation}{section}

\newcommand{\pd}{\partial}
\newcommand{\od}{\mathring{D}}
\newcommand{\dgv}{\delta^{g}_{\xi_{SR1}}}
\newcommand{\dgw}{\delta^{g}_{\xi_{SR2}}}
\newcommand{\nn}{\nonumber}

\def\beq{\begin{equation}}
\def\eeq{\end{equation}}
\def\beqa{\begin{eqnarray}}
\def\eeqa{\end{eqnarray}}

\title{BMS algebra at timelike infinity}
\author{Anupam A H, Athira P V}
\date{April 2019}

\begin{document}

\baselineskip 24pt

\begin{center}

{\LARGE Generalized  BMS Algebra at Timelike Infinity \par}

\end{center}

\vskip .5cm
\medskip

\vspace*{4.0ex}

\baselineskip=18pt

\centerline{\large \rm Anupam A H$^{a}$, Aniket Khairnar$^{b,c}$, Arpan Kundu$^{a}$}

\vspace*{4.0ex}

{\it$^a$The Institute of Mathematical Sciences, Homi Bhabha National Institute(HBNI),\\
\centerline{IV Cross Road, CIT Campus, Taramani, Chennai -- 600113,
Tamil Nadu, India }}


{\it $^b$Department of Physics and Astronomy, University of Mississippi, Oxford-38677, USA.}

{\it $^c$Department of Physics, Indian Institute of Science Education and Research,\\\centerline{ Dr. Homi Bhabha Road, Pashan, Pune - 411008, India.}}

\vspace*{1.0ex}


\vspace*{1.0ex}

\vspace*{1.0ex}
\centerline{\small {\textit {E-mail :}}}  \href{mailto:anupam@imsc.res.in}{ \texttt{anupam@imsc.res.in}}, \href{mailto:akhairna@go.olemiss.edu}{\texttt{akhairna@go.olemiss.edu}},
\href{mailto:akundu@imsc.res.in}{\texttt{akundu@imsc.res.in}}

\vspace*{5.0ex}

\centerline{\bf Abstract} \bigskip
BMS group (and it's various generalizations) at null infinity have been studied extensively in the literature as the symmetry group of asymptotically flat spacetimes.  The intricate relationship between soft theorems and the BMS symmetries have also motivated definition of such asymptotic symmetries to time-like infinity \cite{miguel-bulk-boundary}. Although the vector fields that generate the (generalized) BMS algebra at time-like infinity was defined in the literature, the algebra has not been investigated. In this paper we fill this gap.  We show that the super-translations and vector fields that generate sphere diffeomorphisms close under the modified Lie bracket proposed by Barnich et al. in \cite{barnich}.


\vfill \eject

\baselineskip=18pt

\tableofcontents

\section{Introduction}

The asymptotic properties of spacetime have been of considerable interest since the second half of the twentieth century. One naively expects that the symmetry group of asymptotically flat spacetimes to be the isometries of flat spacetime, i.e., Poincar\'e group. But, in their seminal work  Bondi, Van der Berg, Metzner and Sachs \cite{bondi1,bondi2,bondi3} showed that, one gets an infinite dimensional extension of the Poincare group. This group is known as the BMS group. It is a semi-direct product of the Lorentz group and an infinite dimensional extension of the translation group, namely supertranslations. The BMS group since then has found many applications in areas of gravitational physics from studying various gravity wave solutions of asymptotically flat spacetimes using numerical relativity to cosmology, exact solution techniques, and quantum gravity \cite{bondi4,bms-review,ama}.


The deep underlying conceptual connection between two seemingly independent directions of research on asymptotic symmetries and soft theorems  were unknown until Strominger et al.\cite{strom} showed that Weinberg's soft graviton theorem is equivalent to the conjectured BMS symmetry\cite{strombms}  of quantum gravity $\mathcal{S}$-matrix. This initiated a renewed interest in understanding certain infrared structures of gauge theory and gravity, namely connection between different soft theorems and the asymptotic symmetry groups and their relation to experimentally observable effects  called memory effects \cite{strominger-review}. An extension of the BMS symmetry to include the local conformal Killing vectors (CKVs) on the conformal sphere at null infinity was proposed by Barnich et al. in the context of proposed BMS-CFT correspondence \cite{barnich}. For gravity, a soft graviton theorem was conjectured at the subleading level and proved using modern amplitude techniques \cite{ashokesubleading}\cite{CS}.  It was shown by Strominger et al \cite{stromsublead}, that the subleading soft graviton theorem can be derived from the Ward identities of subgroup of the extended BMS symmetries (called superrotations) as proposed by Barnich et al. In \cite{alok:subleading}, the authors have shown that if one considers a different extension of original BMS group (also known as generalized BMS group), namely instead of CKVs,  the group of smooth diffeomorphism on the conformal sphere at null infinity, the Ward identities resulting from the corresponding charges\cite{alok-new-symmetries} can be shown to be equivalent to the subleading soft graviton theorem. The existence of a theory dependent non universal sub-subleading soft theorem \cite{alok-sub-subleading} from large diffeomorphism symmetry of Einstein's gravity has been proven by Campiglia and Laddha \cite{alok:gauge-gravity}.

In all these analysis, the focus has been on the asymptotic symmetry group at null infinity and its relationship with the soft theorems where the external particles were massless. In soft theorems, the external particles (other than the soft particle), can be massive or massless. To prove the equivalence between asymptotic symmetries and soft theorems when the external states contains massive particles, one needs to include the phase space for massive particles as well. Based on the earlier work on the action of BMS group on massive scalar particle phase space \cite{canonical}, this question was addressed in \cite{alok:massive}. Massive particle geodesics asymptotically reaches to timelike infinity in an asymptotically flat spacetime. In \cite{alok:massive}, the authors considered constant time Euclidean-AdS hypersurface foliations of the Minkowski space. In the limit when the time coordinate in their coordinate system tends to infinity, one reaches near timelike infinity. Since, the boundary of such hypersurfaces resides on the null infinity, one can express the vector fields preserving large time fall off behaviour of Minkowski space, using the generalized BMS vector fields, by use of bulk-boundary Green's functions of standard AdS/CFT dictionary. In this way, one has a natural action of generalized BMS vector fields near timelike infinity, which are intrinsically defined from the perspective of null infinity.

The study of extended BMS charge algebra at null infinity has been extensively studied in \cite{barnich},\cite{barnich2}. Recently the relationship of this algebra at null infinity  with a special class of double soft graviton theorems called consecutive double soft theorems has been explored in  \cite{distler}\cite{anupam}. Double soft graviton theorems are factorization theorems involving two soft gravitons.  
Motivated by these works, our main goal is to understand the generalized BMS charge algebra from the perspective at timelike infinity and its relationship with double soft theorems when the external states are massive. This paper serves as a precursor to this goal. In this work, we are interested in understanding the generalized BMS vector field algebra at timelike infinity with the aim of understanding the generalized BMS charge algebra at timelike infinity in future. We show that there is a closure of generalized BMS vector fields under modified version of Lie bracket as proposed by Barnich et.al. It is important to note that, for similar questions in (2+1) dimensions progress was already made in \cite{3d}.  

 The rest of the paper is organised as follows. Section \ref{rni} deals with the algebra of generalized BMS vector fields at null infinity. In section \ref{rti}, we discuss the asymptotic flatness at timelike infinity and associated generalized BMS vector fields at timelike infinity. We also discuss the constraints on the vector fields and what we mean by supertranslation and the Diff($S^{2}$) vector fields from the perspective of timelike infinity. The need for modified Lie bracket for realising the vector field algebra is also summarized. In section \ref{dti}, we show the algebra between generalized BMS vector fields at timelike infinity and prove that there is a closure of the vector fields. We conclude and address about the further directions in section \ref{cn}.
 

\section{Generalized BMS vector fields}\label{prel}	
\subsection{Generalized BMS vector fields at Null infinity}\label{rni}
We start by reviewing the generalized BMS vector fields and their algebra at null infinity. We discuss the case for future null infinity following \cite{alok:gauge-gravity}, but similar analysis can be done for past null infinity.

The coordinates that are well adapted for describing future null infinity are ($u,r,x^{A}$), where $u=t-r$ is the retarded time, $r$ is the radial coordinate and $x^{A}$ denote the direction along the unit sphere $S^2$. One can reach future null infinity by taking $u=\mathrm{const}$ and $r\rightarrow\infty$ limit. The flat Minkowski metric in these coordinates is given by the line element
\begin{align}
ds^{2}=-du^{2}-2dudr+r^{2}q_{AB}dx^{A}dx^{B}
\end{align}
where $q_{AB}$ is the unit $S^2$ metric. The generalized BMS vector fields can be described as follows. These are vector fields (denoted by $\xi^{a}$) that survive at null infinity and generate residual gauge transformations in the de-Donder gauge (w.r.t to Minkowski metric). Such vector fields obey the wave equation. Additionally they satisfy the asymptotic divergence free condition as given in \cite{alok:subleading}. These two conditions can be written as
\begin{align}\label{wave}
\square\xi^{a}=0\\\label{divfree}
\lim\limits_{r\rightarrow\infty}\nabla_{a}\xi^{a}=0
\end{align}

where $\square,\nabla$ refers to the flat space Laplacian and flat space covariant derivative respectively. In order to understand the structure of the vector fields that satisfy these conditions, one starts with the following ansatz:
\begin{align}\nonumber
\xi^{a}\partial_{a}=\Big(\xi^{(0)u}(u,x^{B})+\mathcal{O}(r^{\epsilon})\Big)&\partial_{u} + \Big(r\xi^{(1)r}(u,x^{B})+\mathcal{O}(r^{0})\Big)\partial_{r}\\
&+\Big(\xi^{(0)A}(u,x^{B})+r^{-1}\xi^{(-1)A}(u,x^{B})+\mathcal{O}(r^{-1-\epsilon})\Big)\partial_{A}
\end{align}

One can find the vector field components by substituting the above ansatz in \ref{wave} and solving them perturbatively in $r$. The details of the computation can be found in \cite{alok:gauge-gravity}. Finally, one gets the generalized BMS vector field as:
\begin{align}\label{nbms}
	\xi=(f+u\alpha)\partial_{u}-r\alpha\partial_{r}+V^{A}\partial_{A}+ \cdots
\end{align}
Here, $f=f(\hat{q})$ is a free scalar function and  $V^{A}=V^{A}(\hat{q})$ is a free vector field which depends on the sphere coordinates $\hat{q}$. Also, $\alpha=\frac{1}{2}D_{A}V^{A}$, where $D_{A}$ is the covariant derivative compatible with $q_{AB}$. The vector fields characterized by the function $f(\hat{q})$ (i.e by setting $V^{A}=0$ in \ref{nbms} ) are  called the supertranslation vector fields. Similarly the vector fields characterized by $V^{A}$ (by setting $f=0$ in \ref{nbms}) are called Diff($S^{2}$) vector fields.
The subleading components in 1/r expansion are also characterized by $f(\hat{q})$ and $V^A(\hat{q})$. The supertranslation and Diff($S^{2}$) vector fields at future null infinity can therefore be written as:
\begin{align}\label{st}
\xi_{f}&=f\partial_{u}\\
\xi_{V}&=u\alpha\partial_{u}-r\alpha\partial_{r}+V^{A}\partial_{A}\label{sr}
\end{align}
 
One can study the algebra of the vector fields by computing the commutator of two variations of the metric w.r.t to the vector fields.
\begin{align}\nonumber
[\delta_{\xi_{1}},\delta_{\xi_{2}}]g_{\mu\nu}&=\delta_{\xi_{1}}\delta_{\xi_{2}}g_{\mu\nu}-\delta_{\xi_{2}}\delta_{\xi_{1}}g_{\mu\nu}\\\nonumber
&=\delta_{\xi_{1}}\mathcal{L}_{\xi_{2}}g_{\mu\nu}-\delta_{\xi_{2}}\mathcal{L}_{\xi_{1}}g_{\mu\nu}\\\nonumber
&=\mathcal{L}_{\xi_{1}}\mathcal{L}_{\xi_{2}}g_{\mu\nu}-\mathcal{L}_{\xi_{2}}\mathcal{L}_{\xi_{1}}g_{\mu\nu}\\
&=\delta_{[\xi_{1},\xi_{2}]}g_{\mu\nu}\label{olbn}
\end{align}
where $[\xi_{1},\xi_{2}]$ denotes the Lie bracket of the vector fields which is defined as,
\begin{align}
[\xi_{1},\xi_{2}]^{a}=\xi_{1}^{b}\partial_{b}\xi_{2}^{a}-\xi_{2}^{b}\partial_{b}\xi_{1}^{a}
\end{align}
Therefore, the generalized vector field algebra at null infinity is found to be,
\begin{align}
[\xi_{f_{1}},\xi_{f_{2}}]=0~~;~~[\xi_{V},\xi_{f}]=\xi_{\widetilde{f}}~~;~~[\xi_{V_{1}},\xi_{V_{2}}]=\xi_{\widetilde{V}}.
\end{align}
Here, $\xi_{f_{1}}$ and $\xi_{f_{2}}$ are two supertranslation vector fields characterized by two functions on the sphere namely, $f_{1}$ and $f_{2}$. Similarly, $\xi_{V_{1}}$ and $\xi_{V_{2}}$ are two Diff($S^{2}$) vector fields characterized by two vector fields on the sphere namely $V_{1}$ and $V_{2}$. Here, $\xi_{\widetilde{f}}$ is another supertranslation vector field characterized by $\widetilde{f}=\mathcal{L}_{V}f-\alpha f$. Also, $\xi_{\widetilde{V}}$ is another Diff($S^{2}$) vector field characterized by $\widetilde{V}^{A}=V_{1}^{B}\partial_{B}V_{2}^{A}-V_{2}^{B}\partial_{B} V_{1}^{A}$. Clearly, supertranslation forms an abelian ideal of the generalized BMS group.
\subsection{Generalized BMS vector fields at Timelike infinity}\label{rti}
Having discussed the algebra of vector fields at null infinity, our main goal in this paper  will be to investigate the algebra at timelike infinity. Following \cite{alok:massive,miguel-bulk-boundary}, we summarize the key ideas that are relevant for our analysis. 
The set of coordinates which we shall be using are the hyperbolic coordinates $(\tau,\rho,\hat{x})$, which are defined in terms of Cartesian coordinates $(t,\vec{x}) $ in the region $t\geq r\equiv \sqrt{\vec{x}\cdot\vec{x}}$ as:
\begin{align}
\tau\coloneqq\sqrt{t^2-r^2}~~;~~~\rho\coloneqq\frac{r}{\sqrt{t^2-r^2}}~~~;~~~\hat{x}=\vec{x}/r
\end{align}
We consider a space of metrics $g_{ab}$ which has an asymptotic expansion in $\tau$ near timelike infinity of the form:
\begin{align}\label{ma}
ds^2=\big(-1+\mathcal{O}(1/\tau)\big)~d\tau^2+\tau^2h_{\alpha\beta}(\tau,\rho,\hat{x})dx^{\alpha}dx^{\beta}
\end{align}
where $h_{\alpha\beta}(\tau,\rho,\hat{x})$ has the following asymptotic expansion (in $\tau$) around timelike infinity
\begin{align}
h_{\alpha\beta}(\tau,\rho,\hat{x})=h_{\alpha\beta}^{(0)}(\rho,\hat{x})+\frac{h_{\alpha\beta}^{(1)}(\rho,\hat{x})}{\tau}+\frac{h_{\alpha\beta}^{(2)}(\rho,\hat{x})}{\tau^{2}}+\cdots.
\end{align}
\\
The notion of asymptotic flatness for metric of this form \ref{ma} at timelike infinity have been addressed in \cite{tanabe},\cite{ugen}. The Minkowski metric (which we denote by $\mathring{g}_{ab}$) belongs to the class of metric \ref{ma} which has only the leading components (in $\tau$) and the hyperboloid components take a particular form. The line element for $\mathring{g}_{ab}$ is written as:
\begin{align}\label{refm}
ds^2=-d\tau^2+\tau^2 \mathring{h}_{\alpha \beta}(\rho,\hat{x})dx^{\alpha}dx^{\beta},
\end{align}
where
\begin{align}\label{eads3}
\mathring{h}_{\alpha\beta}(\rho,\hat{x}) dx^{\alpha}dx^{\beta}\equiv \frac{d\rho^2}{1+\rho^2}+\rho^2q_{AB}dx^{A}dx^{B}.
\end{align}
Here, $q_{AB}$ is the unit metric on 2-sphere. The greek indices $\alpha,\beta, \cdots$ runs over the coordinates on the hyperboloid and the capital Latin indices $A,B,C,\cdots$ runs over the co-ordinates of the 2-sphere.
Here after, we denote the small Latin indices $a,b,c, \cdots$ to denote the four spacetime indices. The Riemann tensor for the above mentioned hyperboloid metric ($\mathring{h}_{\alpha\beta}$) can be written as
\begin{align}\label{hrt}
\mathring{R}_{\alpha\beta\gamma\delta}=\mathring{h}_{\alpha\delta}\mathring{h}_{\beta\gamma}-\mathring{h}_{\alpha\gamma}\mathring{h}_{\beta\delta}~~~~~~;~~~~~\mathring{R}^{\alpha}_{\beta\rho\gamma}=\delta^{\alpha}_{\gamma}\mathring{h}_{\beta\rho}-\delta^{\alpha}_{\rho}\mathring{h}_{\beta\gamma}
\end{align}	
One can reach timelike infinity $i^{+}$ in hyperboloid coordinates by taking $\tau\rightarrow\infty$ limit (or in the Cartesian coordinates $t\rightarrow\infty$ keeping $t\geq r$ ). Similarly, one can reach the part of null infinity where $u>0$ in the hyperboloid coordinates by taking the limit $\tau\rightarrow\infty$, $\rho\rightarrow\infty$, keeping $\frac{\tau}{2\rho}=\mathrm{const}$.

In order to analyze the asymptotic symmetries at timelike infinity $i^{+}$ we suitably adapt the de-Donder gauge in the hyperbolic coordinates. In this gauge, the residual (large) diffeomorphisms are precisely generated by supertranslation and Diff$(S^{2})$ vector fields that smoothly matches with the corresponding BMS vector fields at null infinity.

We consider the following gauge conditions to the metric ansatz\footnote{We are indebted to Miguel Campiglia for suggesting this gauge choice which was a vital input in this work.} \ref{ma}
\begin{align}\label{gaugec}
\mathring{\nabla}_{b}\mathcal{G}^{ab}&=0~~~~~~~\\~~~~~~\mathrm{Tr}(h^{(1)}_{\alpha\beta}(\rho,\hat{x}))&=0\label{trace}
\end{align}
where $\mathcal{G}^{ab}\equiv\sqrt{g}g^{ab}$ and $\mathring{\nabla}_{b}$ refers to the covariant derivative w.r.t to the reference Minkowski metric ($\mathring{g}_{ab}$) in \ref{refm}. One can see that the gauge condition \ref{gaugec} reduces to the usual de-Donder gauge condition when one uses the linearised metric around the Minkowski metric , i.e $g_{ab}\rightarrow \mathring{g}_{ab}+h_{ab}$\footnote{Not to be confused $h_{ab}$ here with the hyperbolid metric defined earlier \ref{eads3}. Here $h_{ab}$  refers to a small perturbation around the Minkowski metric }. It is also important to note that, the trace free condition \ref{trace} of $h^{(1)}_{\alpha\beta}(\rho,\hat{x})$ is taken w.r.t to $h^{(0)}_{\alpha\beta}(\rho,\hat{x})$.

The generalized BMS vector fields at timelike infinity are those that generate the group of diffeomorphisms that preserve the form of the metric \ref{ma} and the gauge conditions \ref{gaugec} and \ref{trace}. To find the structure of such vector fields we start by taking a general ansatz for the vector fields which has an asymptotic expansion (in $\tau$) of the form:
\begin{align}\label{va}
\xi(\tau,\rho,\hat{x})=\Big(\xi^{(0)\tau}(\rho,\hat{x})+\frac{\xi^{(1)\tau}(\rho,\hat{x})}{\tau}+\cdots\Big)\partial_{\tau}+\Big(\xi^{(0)\alpha}(\rho,\hat{x})+\frac{\xi^{(1)\alpha}(\rho,\hat{x})}{\tau}+\cdots\Big)\partial_{\alpha}.
\end{align}
\\
From the form of the metric ansatz given in \ref{ma}, we note that the metric component $g_{\tau\alpha}$ is absent. This imposes the following condition on the vector field:
\begin{align}\label{ssc2}
\mathcal{L}_{\xi}g_{\tau\alpha}=0 ~~~&\Longleftrightarrow ~~~~ \xi^{(1)\alpha}(\rho,\hat{x})=D^{\alpha}\xi^{(0)\tau}(\rho,\hat{x}).
\end{align}
Here $D^{\alpha}$ refers to the covariant derivative w.r.t $h^{(0)}_{\alpha\beta}(\rho,\hat{x})$. Similarly the trace free condition \ref{trace} leads to the following constraint.
\begin{align}
h^{(0)\alpha\beta}\mathcal{L}_{\xi}g_{\alpha\beta}=0~~\mathrm{at} ~\mathcal{O}(\tau^{0})~~~~~ &\Longleftrightarrow \Big(\Delta -3 \Big)\xi^{(0)\tau}(\rho,\hat{x})=0.\label{ssc1}
\end{align}
Here $\Delta$ refers to the Laplacian w.r.t $h^{(0)}_{\alpha\beta}(\rho,\hat{x})$. 
The remaining gauge condition \ref{gaugec} can also be written as
\begin{align}\label{nlg}
g^{ab}\partial_{b}\Big(\ln\Big(\sqrt{\frac{h}{\mathring{h}}}\Big)\Big)+\mathring{\nabla}_{b}g^{ab}=0
\end{align}
The above expression puts the following contraints on the vector fields (details are given in the Appendix-\ref{constraints}): 
\begin{align}
2~D^{(\alpha}\xi^{(0)\beta )}\pd_{\beta}\Big(\ln\Big(\sqrt{\frac{h^{(0)}}{\mathring{h}}}\Big)\Big)~+~ h^{(0) \alpha\beta}\pd_{\beta}D_{\gamma}\xi^{(0)\gamma}~+~2	\mathring{D}_{\beta}D^{(\alpha}\xi^{(0)\beta )}=0,\label{nlc3}\\
D^{\alpha}\xi^{(0)\beta} \mathring{h}_{\alpha\beta}=0.\label{nlc4}
\end{align}
In the above expression $\mathring{D}_{\beta}$ refers to the covariant derivative w.r.t reference hyperboloid metric $\mathring{h}_{\alpha\beta}$. As one can see through the constraints \ref{ssc1}, \ref{nlc3} and \ref{nlc4}, the vector field components (to the leading order in $\tau$) depend upon the hyperboloid metric $h_{\alpha\beta}^{(0)}$ as well as the reference hyperboloid metric $\mathring{h}_{\alpha\beta}$. The dependance on $\mathring{h}_{\alpha\beta}$ arises due to the gauge condition \ref{gaugec} that we have chosen in which divergence is taken w.r.t to the reference metric $\mathring{g}_{ab}$.

In \cite{alok:massive}, Campiglia and Laddha derived the generalized BMS vector fields at timelike infinity as residual gauge transformations (that survive at timelike infinity) of de-Donder gauge around the fixed Minkowski background $\mathring{g}_{ab}$. The conditions that we obtained for the vector fields is more general in the sense that these are the contraints for the vector fields that preserve the form of the metric ansatz\footnote{The fixed Minkowski metric is one of the metric that satisfies the ansatz.} together with the gauge conditions. Inorder to make connection with \cite{alok:massive}, we consider the above constraints \ref{ssc1}, \ref{nlc3} and \ref{nlc4} evaluated at $h_{\alpha\beta}^{(0)}=\mathring{h}_{\alpha\beta}$. Therefore, substituting $h_{\alpha\beta}^{(0)}=\mathring{h}_{\alpha\beta}$ in \ref{ssc1}, \ref{nlc3} and \ref{nlc4} we get,
\begin{align}
\Big( \mathring{\Delta} -3 \Big)\xi^{(0)\tau}(\rho,\hat{x})=0\label{st1}\\
\Big( \mathring{\Delta} -2 \Big)\xi^{(0)\alpha}(\rho,\hat{x})=0\label{sr1}\\
\mathring{D}_{\alpha}\xi^{(0)\alpha}(\rho,\hat{x})=0\label{sr2}
\end{align}  
where $\mathring{\Delta}$ refers to Laplacian w.r.t $\mathring{h}_{\alpha\beta}$. These are the same conditions that the authors arrive in \cite{alok:massive} for the vector fields at timelike infinity. The following boundary conditions are also imposed to make connection with the generalised BMS vector fields at null-infinity.
\begin{align}
\lim\limits_{\rho\rightarrow\infty}\rho^{-1} \xi^{(0)\tau}(\rho,\hat{x})=f(\hat{x}),\\
\lim\limits_{\rho\rightarrow\infty} \xi^{(0)A}(\rho,\hat{x})=V^{A}(\hat{x}).
\end{align}
From the above equations, the leading component of these vector fields can be written in terms of the functions characterizing supertranslation and Diff($S^{2}$) vector field at null infinity
\begin{align}
\xi^{(0)\tau}(\rho,\hat{x})&=\int_{S^{2}}d^{2}\hat{q}~G_{ST}(\rho,\hat{x};\hat{q})f(\hat{q})\equiv f_{\mathcal{H}}(\rho,\hat{x}),\label{nulltimest}
\\
\xi^{(0)\alpha}(\rho,\hat{x})&=\int_{S^{2}}d^{2}\hat{q}~G^{\alpha}_{A}(\rho,\hat{x};\hat{q})V^{A}(\hat{q})\equiv V_{\mathcal{H}}^{\alpha}(\rho,\hat{x}).\label{nulltimesr}
\end{align}	
\\
The Green's functions in turn follows the following constraints:
\begin{align}
(\mathring{\Delta}-3)G_{ST}=0 ~~~~~~~;~~~~~~~~~\lim\limits_{\rho\rightarrow\infty}\rho^{-1} G_{ST}(\rho,\hat{x};\hat{q})=\delta^{(2)}(\hat{x},\hat{q})\label{stgfdef}\\
(\mathring{\Delta}-2)G^{\alpha}_{A}=0 ~~~~;~~~~~\mathring{D}_{\alpha}G^{\alpha}_{A}=0~~;~~\lim\limits_{\rho\rightarrow\infty} G^{A}_{B}(\rho,\hat{x};\hat{q})=\delta^{A}_{B}~\delta^{(2)}(\hat{x},\hat{q})\label{srgfdef}
\end{align}
For detailed expressions of the Green's functions and further discussions one can refer to \cite{miguel-bulk-boundary}.

 In this work, we are primarily interested in the algebra of the generalized BMS vector fields w.r.t reference Minkowski metric ($\mathring{g}_{ab}$). Therefore the supertranslation and Diff($S^{2}$) vector fields to leading order at timelike infinity are given by
\begin{align}
\xi_{ST}=f_{\mathcal{H}}(\rho,\hat{x})\partial_{\tau}\\
\xi_{SR}=V_{\mathcal{H}}^{\alpha}(\rho,\hat{x})\partial_{\alpha}
\end{align} 

One can verify that variation w.r.t. the supertranslation vector field does not alter the leading order (in $\tau$) structure of \ref{ma} (and hence \ref{refm}) but the variation under Diff($S^{2}$) vector field does. This can be seen from evaluating the Lie derivative of the metric w.r.t supertranslation/Diff($S^{2}$) vector field.
\begin{align}\label{stc}
\mathcal{L}_{\xi_{ST}}\mathring{g}_{\tau\tau}=0~~~~~;~~~~\mathcal{L}_{\xi_{ST}}\mathring{g}_{\alpha\beta}=\mathcal{O}(\tau)\\
\mathcal{L}_{\xi_{SR}}\mathring{g}_{\tau\tau}=0~~~~;~~~\mathcal{L}_{\xi_{SR}}\mathring{g}_{\alpha\beta}=\mathcal{O}(\tau^{2}) 	\label{src}
\end{align}
One can clearly see that the Diff($S^{2}$) vector field changes the hyperboloid components of the metric at order $\tau^{2}$. The relevance of the above mentioned point will become clear in further sections where we verify the algebra of the vector fields.

Our main interest in this paper is to understand that whether the supertranslation and Diff($S^{2}$) vector fields defined above form a closed algebra at time-like infinity. A naive attempt to study these algebra will be to compute the ordinary Lie bracket (as we have done for the null infinity case) of the vector fields and check whether the resulting vector field satisfies the constraint \ref{st1} (in case for supertranslation), \ref{sr1} and \ref{sr2} (in case for Diff($S^{2}$)). However, as is well known in the literature \cite{barnich},\cite{barnich2}, the correct definition of Lie bracket in the case of asymptotic symmetries is more intricate. This can be explained as follows.

Usually, one studies the vector field algebra by considering the commutator of two variations of the vector fields on the metric. An important point to be noted here is the fact that, the vector fields themselves are metric dependant\footnote{This was not the case  at null infinity, where the generalized BMS vector fields were metric independant.}. This can be seen from the defining equations for the vector field \ref{st1}, \ref{sr1} and \ref{sr2}, which tells us that the vector fields depend upon the hyperboloid metric $\mathring{h}_{\alpha\beta}$ through covariant derivative and Laplacian. Therefore, performing the second variation will affect both the first variation as well as the metric. This can be seen as
\begin{align}\nonumber
[\delta_{\xi_{1}(g)},\delta_{\xi_{2}(g)}]g_{\mu\nu}&=\delta_{\xi_{1}(g)}\delta_{\xi_{2}(g)}g_{\mu\nu}-\delta_{\xi_{2}(g)}\delta_{\xi_{1}(g)}g_{\mu\nu}\\\nonumber
&=\delta_{\xi_{1}(g)}\mathcal{L}_{\xi_{2}(g)}g_{\mu\nu}-\delta_{\xi_{2}(g)}\mathcal{L}_{\xi_{1}(g)}g_{\mu\nu}\\\nonumber
&=\mathcal{L}_{\xi_{1}(g)}\mathcal{L}_{\xi_{2}(g)}g_{\mu\nu}-\mathcal{L}_{\delta_{\xi_{1}}^{g}\xi_{2}(g)}g_{\mu\nu}-\mathcal{L}_{\xi_{2}(g)}\mathcal{L}_{\xi_{1}(g)}g_{\mu\nu}+\mathcal{L}_{\delta_{\xi_{2}}^{g}\xi_{1}(g)}g_{\mu\nu}\nonumber\\
&=\delta_{[\xi_{1}(g),\xi_{2}(g)]}g_{\mu\nu}~-~\big(\delta_{\delta_{\xi_{1}}^{g}\xi_{2}(g)}-\delta_{\delta_{\xi_{2}}^{g}\xi_{1}(g)}\big)g_{\mu\nu}\nonumber\\
&=\delta_{\big([\xi_{1}(g),\xi_{2}(g)]-\delta_{\xi_{1}}^{g}\xi_{2}(g)+\delta_{\xi_{2}}^{g}\xi_{1}(g)\big)}g_{\mu\nu}
\end{align}
\\
As one can see, this is different from \ref{olbn}. The first term in the above expression is the ordinary Lie bracket which is same as the one we encountered in the null infinity case. The extra term $\delta_{\xi_{1}}^{g}\xi_{2}(g)$ captures the variation on the  vector field $\xi_{2}(g)$  due to the action of the vector field $\xi_{1}(g)$ on the metric. Hence, in order to realise the algebra of the vector fields at timelike infinity one needs to take into account such terms. One defines the modified Lie bracket for realising the BMS vector fields algebra as
\begin{align}\label{modified-def}
[\xi_{1},\xi_{2}]^{a}_{M}\equiv[\xi_{1},\xi_{2}]^{a}-\delta^{g}_{\xi_{1}}\xi^{a}_{2}+\delta^{g}_{\xi_{2}}\xi^{a}_{1}
\end{align}
where $\delta^{g}_{\xi_{1}}\xi^{a}_{2}$ denotes the change in $\xi^{a}_{2}$ due to the variation in the metric induced by $\xi_{1}$. The exact computation of these terms will be shown in the next section.

We end this section by emphasising the difference between the two set of constraints we have derived for the vector fields. The first set of constraints (equations \ref{ssc1}, \ref{nlc3} and \ref{nlc4}) are the defining equations for the vector fields that preserve the gauge conditions and the metric ansatz \ref{ma}. The second set of constraints (equations \ref{st1}, \ref{sr1} and \ref{sr2}) are the conditions on the vector fields when,	 one chooses a particular metric from the metric ansatz, i.e, the reference Minkowski metric \ref{refm}. 

\section{Generalised BMS vector field algebra at timelike infinity}\label{dti}
In this section, we show the closure of the generalized BMS vector fields at timelike infinity using the modified Lie-bracket. We first consider the algebra between two supertranslations and then, in the next sub-section, we look at the algebra between a supertranslation and Diff($S^{2}$) vector field. Finally, we would be considering the algebra between two Diff($S^{2}$) vector fields. In each case, we find a similar result like one gets for the algebra for generalized BMS vector fields at null infinity.
\subsection{Algebra between two Supertranslations}\label{2ST}
We start with the case of two supertranslations. Consider two supertranslation
vector fields:
\begin{align}\label{two-supertranslation}
	\xi_{ST1}=f_{\mathcal{H}}(\rho,\hat{x})\partial_{\tau}\\
	\xi_{ST2}=g_{\mathcal{H}}(\rho,\hat{x})\partial_{\tau}\label{two-supertranslation2}
\end{align}
where, $f_{\mathcal{H}}$ and $g_{\mathcal{H}}$ is defined as follows:
\begin{align}
	f_{\mathcal{H}}(\rho,\hat{x})=\int d^{2}\hat{q}_{1}~G_{ST1}(\rho,\hat{x};\hat{q_{1}})f(\hat{q}_{1})\\
	g_{\mathcal{H}}(\rho,\hat{x})=\int d^{2}\hat{q}_{2}~G_{ST2}(\rho,\hat{x};\hat{q_{2}})g(\hat{q}_{2})
\end{align}
Here, $G_{ST1}$ and $G_{ST2}$ is the same Green's function satisfying the constraints \ref{stgfdef}. In order to compute the algebra of two supertranslation vectors, we evaluate the modified Lie bracket as defined in \ref{modified-def}. We expect an algebra similar to the case of null infinity, where the supertranslation vector fields commute.

The modified Lie bracket is written as:
\begin{align}\label{modified-two-supertranslation}
[\xi_{ST1},\xi_{ST2}]^{a}_{M}=[\xi_{ST1},\xi_{ST2}]^{a}-\delta^{g}_{\xi_{ST1}}\xi^{a}_{ST2}+\delta^{g}_{\xi_{ST2}}\xi^{a}_{ST1}
\end{align}
As we have explained in the previous section, supertranslation vector fields do not change
the Minkowski metric at the leading order in $\tau$. This can be seen from \ref{stc}. Hence, the terms $\delta^{g}_{\xi_{ST1}}\xi^{a}_{ST2}$ and $\delta^{g}_{\xi_{ST2}}\xi^{a}_{ST1}$ do not contribute at  timelike infinity. Consequently, the above expression of modified Lie bracket reduces to the ordinary Lie bracket, namely:
\begin{align}
[\xi_{ST1},\xi_{ST2}]^{a}_{M}=[\xi_{ST1},\xi_{ST2}]^{a}
\end{align}
Now, using the expressions of the vector fields  \ref{two-supertranslation} and \ref{two-supertranslation2}, it is then easy to see that ordinary Lie bracket also vanishes. Hence we finally get
\begin{align}
[\xi_{ST1},\xi_{ST2}]^{a}_{M}=0.
\end{align}
This matches with the case of null infinity. We see, similar to null infinity, supertranslations form an abelian ideal.

%

\subsection{Algebra between a Supertranslation and a Diff($S^{2}$) vector field}\label{STSR}
We now consider the modified Lie
bracket between a supertranslation and a Diff($S^{2}$) vector field. i.e: 
\begin{align}\label{2super-fun}
	\xi_{ST}=f_{\mathcal{H}}(\rho,\hat{x})\partial_{\tau}\\
	\xi_{SR}=V^{\alpha}_{\mathcal{H}}(\rho,\hat{x})\partial_{\alpha}\label{2super-fun1}
\end{align}
where, $f_{\mathcal{H}}$ and $V^{\alpha}_{\mathcal{H}}$ are already defined in \ref{nulltimest} and \ref{nulltimesr}, and they satisfy:
\begin{align}\label{stsr1}
	\mathring{\Delta} f_{\mathcal{H}}(\rho,\hat{x})=3f_{\mathcal{H}}(\rho,\hat{x})~~~~;~~~\mathring{D}_{\alpha}V^{\alpha}_{\mathcal{H}}(\rho,\hat{x})=0~~~;~~~\mathring{\Delta} V^{\alpha}_{\mathcal{H}}(\rho,\hat{x})=2V^{\alpha}_{\mathcal{H}}(\rho,\hat{x})
\end{align}

From the equations above it is clear that $f_{\mathcal{H}}$ and $V^{\alpha}_{\mathcal{H}}$ depend upon the metric $\mathring{h}_{\alpha\beta}$ (through covariant derivative $\mathring{D}_{\alpha}$ and Laplacian $\mathring{\Delta}$).

Using \ref{modified-def} the modified Lie bracket of supertranslation and Diff($S^{2}$) vector field can be written as:
\begin{align}\label{stsrm}
[\xi_{ST},\xi_{SR}]^{a}_{M}=[\xi_{ST},\xi_{SR}]^{a}-\delta^{g}_{\xi_{ST}}\xi^{a}_{SR}+\delta^{g}_{\xi_{SR}}\xi^{a}_{ST}
\end{align}
 As explained in the beginning of this section, the  Diff($S^{2}$) vector field depends upon $\mathring{h}_{\alpha\beta}$,  and $\delta^{g}_{\xi_{ST}}\xi^{a}_{SR}$ represents the variation in $\xi^{a}_{SR}$ due to the change in the metric induced by the supertranslation vector field $\xi_{ST}$. But, we already saw in the previous section that, the supertranslation does not alter the Minkowski metric to the leading order \ref{stc} and hence, does not alter $\mathring{h}_{\alpha\beta}$. Therefore, the term $\delta^{g}_{\xi_{ST}}\xi^{a}_{SR}$ in the above expression vanishes and the modified Lie bracket becomes
\begin{align}\label{stsrm1}
[\xi_{ST},\xi_{SR}]^{a}_{M}=[\xi_{ST},\xi_{SR}]^{a}+\delta^{g}_{\xi_{SR}}\xi^{a}_{ST}
\end{align}
From the definitions of the vector fields given in \ref{2super-fun}, \ref{2super-fun1}, it is clear that, only the $\tau$ component contributes to the above expression of modified Lie bracket. For the null infinity case, the algebra of one supertranslation and one Diff($S^{2}$) vector field gives another supertranslation. Hence, it is natural to expect that a similar algebra holds at timelike infinity. Namely, the modified Lie bracket \ref{stsrm1} gives us another supertranslation. In order to verify this, we check whether the conditions on a supertranslation vector field hold for the modified Lie bracket, \textit{i.e.} we check whether
\begin{align}\label{st-sr-main}
(\mathring{\Delta}-3)[\xi_{ST},\xi_{SR}]^{\tau}_{M}\stackrel{?}{=}0.
\end{align}
Or, equivalently,
\begin{align}\label{modified-st-sr-nonzero}
(\mathring{\Delta}-3)[\xi_{ST},\xi_{SR}]^{\tau} + (\mathring{\Delta}-3)\delta^{g}_{\xi_{SR}}\xi^{\tau}_{ST}\stackrel{?}{=}0
\end{align}
In the rest of this section, we show that this is indeed true.
We start with the contribution from the ordinary Lie bracket term.
\begin{align}\label{stsrd1}
(\mathring{\Delta}-3)[\xi_{ST},\xi_{SR}]^{\tau}=-(\mathring{\Delta}-3)\Big[V^{\alpha}_{\mathcal{H}}\mathring{D}_{\alpha}f_{\mathcal{H}}\Big]
\end{align}
Using the properties of $V^{\alpha}_{\mathcal{H}}$ and $f_{\mathcal{H}}$ given in \ref{stsr1}, the r.h.s of the above expression finally becomes (Details of the calculation are given in Appendix-\ref{st-sr-ord-calc}):
\begin{align}\label{stsre6}
(\mathring{\Delta}-3)[\xi_{ST},\xi_{SR}]^{\tau}=-2\mathring{D}^{\beta}V^{\alpha}_{\mathcal{H}}\mathring{D}_{\beta} \mathring{D}_{\alpha}f_{\mathcal{H}}
\end{align}
We now proceed to evaluate the second term in \ref{modified-st-sr-nonzero}. As we have mentioned in the previous section, the Diff($S^{2}$) vector field changes the Minkowski metric at the leading order. It can be easily seen that, under the Lie derivative action of the Diff($S^{2}$) vector field, the hyperboloid components of the reference Minkowski metric is shifted, i.e
\begin{align}
\mathcal{L}_{\xi_{SR}}\mathring{g}_{\alpha\beta}=\tau^{2}\big( \mathring{D}_{\alpha}\xi_{SR \beta}+\mathring{D}_{\beta}\xi_{SR\alpha}\big)
\end{align}
where $\mathring{D}$ refers to the covariant derivative w.r.t. to reference hyperboloid metric $\mathring{h}_{\alpha\beta}$. Thereby, the gauge condition on the supertranslation vector fields shift to 
\begin{align}
( \Delta -3 )\xi_{ST}^{\tau}=0,
\end{align}
where, $\Delta$ refers to the Laplacian w.r.t. to shifted hyperboloid $h_{\alpha\beta}^{(0)}=\mathring{h}_{\alpha\beta}+\mathcal{L}_{\xi_{SR}}\mathring{h}_{\alpha\beta}$. This indicates that the change in the vector field $\xi_{ST}$ due to the change in the metric induced by $\xi_{SR}$ is reflected in the variation of the Laplacian induced by $\xi_{SR}$. Therefore, the second term in  \ref{modified-st-sr-nonzero} can be evaluated as: 
\begin{align}\label{stsre4}
(\mathring{\Delta}-3)\delta^{g}_{\xi_{SR}}\xi^{\tau}_{ST}&=\delta^{g}_{\xi_{SR}}\Big((\mathring{\Delta}-3)\xi^{\tau}_{ST}\Big)-\delta^{g}_{\xi_{SR}}\Big(\mathring{\Delta}-3\Big)\xi^{\tau}_{ST}\nonumber\\&
=-\delta^{g}_{\xi_{SR}}\Big(\mathring{\Delta}-3\Big)\xi^{\tau}_{ST}
\end{align}
In going from first line to the second in the above expression we have used  the fact $(\mathring{\Delta}-3)\xi^{\tau}_{ST}=0$. One can evaluate r.h.s of   \ref{stsre4} to (Details of this calculation are given in Appendix-\ref{st-sr-mod-calc}):
\begin{align}\label{stsre5}
(\mathring{\Delta}-3)\delta^{g}_{\xi_{SR}}\xi^{\tau}_{ST}=2\mathring{D}^{\beta}V^{\alpha}_{\mathcal{H}}\mathring{D}_{\beta} \mathring{D}_{\alpha}f_{\mathcal{H}}.
\end{align}
Therefore, summing \ref{stsre6} and \ref{stsre5} we finally get:
\begin{align}
(\mathring{\Delta}-3)[\xi_{ST},\xi_{SR}]^{\tau} + (\mathring{\Delta}-3)\delta^{g}_{\xi_{SR}}\xi^{\tau}_{ST}=0.
\end{align}
This shows that the modified Lie bracket of a supertranslation and a Diff($S^{2}$) vector field is indeed another supertranslation.   
\bigskip
\subsection{Algebra between two Diff($S^{2}$) vector fields}\label{2SR}
We now proceed  to compute the algebra of two Diff($S^{2}$) vector fields at $i^{+}$. The Diff($S^{2}$) vector fields at $i^{+}$ are 
\begin{align}\label{twosre}
	\xi_{SR1}=V^{\alpha}_{\mathcal{H}}(\rho,\hat{x})\partial_{\alpha},\\
	\xi_{SR2}=W^{\alpha}_{\mathcal{H}}(\rho,\hat{x})\partial_{\alpha},\label{twosre1}
\end{align}
where, $V^{\alpha}_{\mathcal{H}}$ and $W^{\alpha}_{\mathcal{H}}$ are defined as in \ref{nulltimesr}. Therefore, we can write:
\begin{align}
	V^{\alpha}_{\mathcal{H}}=\int d^{2}\hat{q}_{1}~G^{\alpha}_{A}(\rho,\hat{x};\hat{q_{1}})V^{A}(\hat{q}_{1}),\\
	W^{\alpha}_{\mathcal{H}}=\int d^{2}\hat{q}_{2}~G^{\alpha}_{B}(\rho,\hat{x};\hat{q_{2}})W^{B}(\hat{q}_{2}),
\end{align}
where $V^{A}(\hat{q}_{1}), ~W^{B}(\hat{q}_{2})$ are two vector fields on the $2-$sphere at $\mathcal{I}^{+}$. The vector fields $V^{\alpha}_{\mathcal{H}},~W^{\alpha}_{\mathcal{H}}$ follow the constraints \ref{srgfdef}.
\begin{align}\label{pp1}
	\mathring{D}_{\alpha}V^{\alpha}_{\mathcal{H}}(\rho,\hat{x})=0~~;~~~\mathring{\Delta} V^{\alpha}_{\mathcal{H}}(\rho,\hat{x})=2V^{\alpha}_{\mathcal{H}}(\rho,\hat{x})\\
	\mathring{D}_{\alpha}W^{\alpha}_{\mathcal{H}}(\rho,\hat{x})=0~~;~~~\mathring{\Delta} W^{\alpha}_{\mathcal{H}}(\rho,\hat{x})=2W^{\alpha}_{\mathcal{H}}(\rho,\hat{x})
\end{align}
In order to understand the algebra between two Diff($S^{2}$) vector fields, we evaluate the modified Lie bracket  \textit{i.e.} 
\begin{align}\label{twosrm}
[\xi_{SR1},\xi_{SR2}]^{a}_{M}=[\xi_{SR1},\xi_{SR2}]^{a}-\delta^{g}_{\xi_{SR1}}\xi^{a}_{SR2}+\delta^{g}_{\xi_{SR2}}\xi^{a}_{SR1}.
\end{align}
It is easy to see that, from the form of the vector fields $\xi_{SR1},~\xi_{SR2}$ given in \ref{twosre}, \ref{twosre1} the $\tau$ component of the modified Lie bracket vanishes and only the hyperboloid component exists. Therefore, we need to evaluate 
\begin{align}\label{twosrm1}
[\xi_{SR1},\xi_{SR2}]^{\alpha}_{M}=[\xi_{SR1},\xi_{SR2}]^{\alpha}-\delta^{g}_{\xi_{SR1}}\xi^{\alpha}_{SR2}+\delta^{g}_{\xi_{SR2}}\xi^{\alpha}_{SR1},
\end{align}
where, $\alpha$ runs over the hyperboloid components only.
At null infinity we have already seen that, the Lie bracket of two Diff($S^{2}$) vector fields is another Diff($S^{2}$) vector field. We expect similar result to hold at timelike infinity. Therefore, we want to check whether the vector field that one gets from the modified Lie bracket obeys the constraints
\begin{align}
\mathring{D}_{\alpha}[\xi_{SR1},\xi_{SR2}]^{\alpha}_{M}\stackrel{?}{=}0\label{twosrdiv}\\
(\mathring{\Delta}-2)[\xi_{SR1},\xi_{SR2}]^{\alpha}_{M}\stackrel{?}{=}0\label{twosrdelta}
\end{align}
Here, written explicitly in terms of expression of modified Lie bracket the above expressions are equivalent to
\begin{align}
\mathring{D}_{\alpha}[\xi_{SR1},\xi_{SR2}]^{\alpha}-\mathring{D}_{\alpha}\delta^{g}_{\xi_{SR1}}\xi_{SR2}^{\alpha}+\mathring{D}_{\alpha}\delta^{g}_{\xi_{SR2}}\xi_{SR1}^{\alpha}&\stackrel{?}{=}0\label{c2}\\
\label{c1}
\Big( \mathring{\Delta} -2 \Big)[\xi_{SR1},\xi_{SR2}]^{\alpha}-\Big( \mathring{\Delta} -2 \Big)\delta^{g}_{\xi_{SR1}}\xi_{SR2}^{\alpha}+\Big( \mathring{\Delta} -2 \Big)\delta^{g}_{\xi_{SR2}}\xi_{SR1}^{\alpha}&\stackrel{?}{=}0.
\end{align}
%
We start with the verfication of \ref{c2}. The first term in the l.h.s of \ref{c2} vanishes. This can be shown as
\begin{align}\nonumber
\mathring{D}_{\alpha}[\xi_{SR1},\xi_{SR2}]^{\alpha}&=V_{\mathcal{H}}^{\beta}\mathring{D}_{\alpha}\mathring{D}_{\beta}W_{\mathcal{H}}^{\alpha}-W_{\mathcal{H}}^{\beta}\mathring{D}_{\alpha}\mathring{D}_{\beta}V_{\mathcal{H}}^{\alpha}
\\&=\mathring{R}^{\alpha}_{\rho\alpha\beta}V_{\mathcal{H}}^{\beta}W_{\mathcal{H}}^{\rho}-\mathring{R}^{\alpha}_{\rho\alpha\beta}W_{\mathcal{H}}^{\beta}V_{\mathcal{H}}^{\rho}\nonumber\\
&=-2\mathring{h}_{\rho\beta}\Big(V_{\mathcal{H}}^{\beta}W_{\mathcal{H}}^{\rho}-W_{\mathcal{H}}^{\beta}V_{\mathcal{H}}^{\rho}\Big)\nonumber\\
&=0
\end{align}
In going from the first line to the second we used the divergence free condition of the Diff($S^{2}$) vector fields. We now proceed to evaluate the contribution from the modification terms (the last two terms in \ref{twosrm1}) in the modified Lie bracket. In the earlier section \ref{STSR}, we showed that the change in the supertranslation vector field due to the change in the metric induced by the Diff($S^{2}$) vector field was reflected in the variation of the Laplacian in \ref{st1}. But the situation is more intricate for the case of Diff($S^{2}$) vector field. The gauge conditions \ref{sr1} and \ref{sr2} for one of the Diff($S^{2}$) vector field (say $\xi_{SR1}$) now shift to \ref{nlc3} and \ref{nlc4} respectively where $h^{(0)}_{\alpha\beta}(\rho,\hat{x})$ will be now defined by $h^{(0)}_{\alpha\beta}=\mathring{h}_{\alpha\beta}+\mathcal{L}_{\xi_{SR2}}\mathring{h}_{\alpha\beta}$,where $\xi_{SR2}$ is another Diff($S^{2}$) vector field. Keeping this in mind, inorder to evaluate the last two terms in the l.h.s of \ref{c2}, we use the residual gauge condition \ref{nlc4}, which is one of the defining condition for the Diff$(S^{2})$ vector field for an arbitrary $h^{(0)}_{\alpha\beta}(\rho,\hat{x})$. We vary this gauge condition w.r.t. another  Diff$(S^{2})$ vector field and finally evaluate the expression at $h^{(0)}_{\alpha\beta}(\rho,\hat{x})=\mathring{h}_{\alpha\beta}(\rho,\hat{x})$. We demonstrate this in detail further in this section.
\\
We start with the gauge condition \ref{nlc4} for an arbitrary $h^{(0)}_{\alpha\beta}$ 
\begin{align}
D^{\alpha}\xi^{(0)\beta} \mathring{h}_{\alpha\beta}=0.
\end{align}
Under variation w.r.t. to the Diff($S^{2}$) vector field $\xi_{\mathrm{SR1}}$, the above condition becomes
\begin{align}
\delta^{g}_{\xi_{SR1}}\Big(D^{\alpha}\xi^{\beta} \mathring{h}_{\alpha\beta}\Big)=\delta^{g}_{\xi_{SR1}}\Big(D^{ \alpha}\Big)\xi^{\beta}\mathring{h}_{\alpha\beta}+D^{ \alpha}\delta^{g}_{\xi_{SR1}}\xi^{\beta}\mathring{h}_{\alpha\beta}=0.
\end{align}
It is important to note that, the variation is not taken on the reference metric $\mathring{h}_{\alpha\beta}$. 
Hence, the above expression can be written as
\begin{align}
D^{ \alpha}\delta^{g}_{\xi_{SR1}}\xi^{\beta}\mathring{h}_{\alpha\beta}=-\delta^{g}_{\xi_{SR1}}\Big(D^{\alpha}\Big)\xi^{\beta}\mathring{h}_{\alpha\beta}.
\end{align}
In order to compute $\mathring{D}_{\alpha}\delta^{g}_{\xi_{SR1}}\xi_{SR2}^{\alpha}$ in \ref{c2}, we evaluate the above expression at $h^{(0)}_{\alpha\beta}=\mathring{h}_{\alpha\beta}$, then
\begin{align}\label{div1}
\mathring{D}^{ \alpha}\delta^{g}_{\xi_{SR1}}\xi^{\beta}_{SR2}\mathring{h}_{\alpha\beta}=-\delta^{g}_{\xi_{SR1}}\Big(\mathring{D}^{\alpha}\Big)\xi^{\beta}_{SR2}\mathring{h}_{\alpha\beta}
\end{align}
The r.h.s of the above expression can be evaluated as
\begin{align}\nonumber
\delta^{g}_{\xi_{SR1}}\Big(\mathring{D}^{\alpha}\Big)\xi^{\beta}_{SR2}\mathring{h}_{\alpha\beta}&=\delta^{g}_{\xi_{SR1}}\Big(\mathring{h}^{\alpha\gamma}\mathring{D}_{\gamma}\Big)\xi^{\beta}_{SR2}\mathring{h}_{\alpha\beta}\\\nonumber
&=\Big(\delta^{g}_{\xi_{SR1}}\mathring{h}^{\alpha\gamma}\Big)\mathring{D}_{\gamma}\xi^{\beta}_{SR2}\mathring{h}_{\alpha\beta}+\mathring{h}^{\alpha\gamma}\Big(\delta^{g}_{\xi_{SR1}}\mathring{D}_{\gamma}\Big)\xi^{\beta}_{SR2}\mathring{h}_{\alpha\beta}\\\nonumber
&=\Big(\mathring{D}^{\alpha}V_{\mathcal{H}}^{\gamma}+\mathring{D}^{\gamma}V_{\mathcal{H}}^{\alpha}\Big)\mathring{D}_{\gamma}W_{\mathcal{H}}^{\beta}\mathring{h}_{\alpha\beta}+\Big(\delta^{g}_{\xi_{SR1}}\mathring{\Gamma}_{\gamma\rho}^{\gamma}\Big)W_{\mathcal{H}}^{\rho}\mathring{h}_{\alpha\beta}\\\nonumber
&=\Big(\mathring{D}^{\alpha}V_{\mathcal{H}}^{\gamma}+\mathring{D}^{\gamma}V_{\mathcal{H}}^{\alpha}\Big)\mathring{D}_{\gamma}W_{\mathcal{H}}^{\beta}\mathring{h}_{\alpha\beta}+0\\
&=\Big(\mathring{D}^{\alpha}V_{\mathcal{H}}^{\gamma}+\mathring{D}^{\gamma}V_{\mathcal{H}}^{\alpha}\Big)\mathring{D}_{\gamma}W_{\mathcal{H}}^{\beta}\mathring{h}_{\alpha\beta}\label{sr5}
\end{align}
In evaluating the above expression, we have used the fact that $\delta^{g}_{\xi_{SR1}}\mathring{\Gamma}_{\gamma\rho}^{\gamma}=0$. This can be easily seen from \ref{gamma-f}. Therefore, \ref{div1} becomes
\begin{align}\label{ch1}
\mathring{D}_{ \alpha}\delta^{g}_{\xi_{SR1}}\xi^{\alpha}_{SR2}=-\mathring{D}_{\gamma}W_{\alpha}\Big(\mathring{D}^{\alpha}V^{\gamma}+\mathring{D}^{\gamma}V^{\alpha}\Big).
\end{align}
Similarly, the last term in \ref{c2} \textit{i.e.}  $\mathring{D}_{\alpha}\delta^{g}_{\xi_{SR2}}\xi_{SR1}^{\alpha}$ can be evaluated as
\begin{align}\label{ch2}
\mathring{D}_{\alpha}\delta^{g}_{\xi_{SR2}}\xi_{SR1}^{\alpha}=-\Big(\mathring{D}^{\alpha}W^{\gamma}+\mathring{D}^{\gamma}W^{\alpha}\Big)\mathring{D}_{\gamma}V_{\alpha}.
\end{align}
Therefore, 
\begin{align}
\mathring{D}_{ \alpha}\delta^{g}_{\xi_{SR1}}\xi^{\alpha}_{SR2}-\mathring{D}_{\alpha}\delta^{g}_{\xi_{SR2}}\xi_{SR1}^{\alpha}=0.
\end{align}
Hence, the divergence of the modified terms sums to zero, thereby verifying one of the conditions for a Diff($S^{2}$) vector field. \textit{i.e.}
\begin{align}
\mathring{D}_{\alpha}[\xi_{SR1},\xi_{SR2}]^{\alpha}_{M}=0.
\end{align}
Now, one needs to verify \ref{c1}.  We start by evaluating the first term in \ref{c1}.
\begin{align}\label{olb1}
\Big( \mathring{\Delta} -2 \Big)[\xi_{SR1},\xi_{SR2}]^{\alpha}&=2\Big(\mathring{D}^{\lambda}V^{\beta}\mathring{D}_{\lambda}\mathring{D}_{\beta}W^{\alpha}-\mathring{D}^{\lambda}W^{\beta}\mathring{D}_{\lambda}\mathring{D}_{\beta}V^{\alpha}\Big)\nonumber\\
&~~~~~~~~~~~~~~~~~~~+\Big(V^{\beta}\mathring{\Delta}~\mathring{D}_{\beta}W^{\alpha}-W^{\beta}\mathring{\Delta}~\mathring{D}_{\beta}V^{\alpha}\Big)
\end{align}
The last two terms can be simplified more using \ref{sr1} and \ref{sr2} and using the identity
\begin{align}
[\mathring{\Delta},\mathring{D}_{\alpha}]T^{a_{1}..a_{n}}&=2\Big(\delta^{a_{1}}_{\alpha}\mathring{D}_{\rho}T^{\rho a_{2}..a_{n}}+..\delta^{a_{n}}_{\alpha}\mathring{D}_{\rho}T^{a_{1}a_{2}..\rho}\Big)\nonumber\\&~~~~~~~~~~~
-2\Big(\mathring{D}^{a_{1}}T_{\alpha}^{a_{2}..a_{n}}+...\mathring{D}^{a_{n}}T^{a_{1}..a_{n-1}}_{\alpha}+\mathring{D}_{\alpha}T^{a_{1}...a_{n}}\Big).
\end{align}
The above expression can be derived using the Riemann tensor of the hyperboloid metric \ref{hrt} and $T^{a_{1}..a_{n}}$ is a arbitrary tensor on the hyperboloid. Therefore, \ref{olb1} finally evaluates to
\begin{align}\nonumber
\Big( \mathring{\Delta} -2 \Big)[\xi_{SR1},\xi_{SR2}]^{\alpha}&=2\Big(\mathring{D}^{\lambda}V^{\beta}\mathring{D}_{\lambda}\mathring{D}_{\beta}W^{\alpha}-\mathring{D}^{\lambda}W^{\beta}\mathring{D}_{\lambda}\mathring{D}_{\beta}V^{\alpha}\Big)\\
&~~~~~~~~~~~~~~~~~~~-2\Big(V^{\beta}\mathring{D}^{\alpha}W_{\beta}-W^{\beta}\mathring{D}^{\alpha}V_{\beta}\Big)\label{udelta}
\end{align}
To evaluate the last two terms in \ref{c2}, we proceed similarly as we have done earlier for the verification of divergence free condition. We use residual gauge condition \ref{nlc3} to evaluate the last two terms. The details of the calculation is given in the Appendix \ref{twosrdeltaa2} . Finally, we get
\begin{align}\nonumber
\Big(\mathring{\Delta}-2\Big)\dgv  \xi_{SR2}^{\alpha}&-\Big(\mathring{\Delta}-2\Big)\dgw \xi_{SR1}^{\alpha}\\
&=2\Big(\od^{\beta}V^{\gamma}\od_{\beta}\od_{\gamma}W^{\alpha}-\od^{\beta}W^{\gamma}\od_{\beta}\od_{\gamma}V^{\alpha}\Big)\nonumber\\
&~~~~~~~~~~~~~~~~~~~~~~~~-2\Big(V_{\gamma}\od^{\alpha}W^{\gamma}-W_{\gamma}\mathring{D}^{\alpha}V^{\gamma}\Big)\label{mfinal}.
\end{align}\\
Therefore, substituting \ref{udelta} and \ref{mfinal} in \ref{c1}, we get
\begin{align}
\Big(\mathring{ \Delta} -2 \Big)[\xi_{SR1},\xi_{SR2}]^{\alpha}-\Big( \mathring{\Delta} -2 \Big)\delta^{g}_{\xi_{SR1}}\xi_{SR2}^{\alpha}+\Big( \mathring{\Delta} -2 \Big)\delta^{g}_{\xi_{SR2}}\xi_{SR1}^{\alpha}=0.
\end{align}
\\
Hence, we have a closure of Diff($S^{2}$) vector field at timelike infinity similar to the case of null infinity.
In all the three cases, the desired relations are satisfied and hence we show  that the BMS vector field algebra closes under the modified Lie bracket.

\section{Conclusions}\label{cn}

In this paper, we showed the closure of generalized BMS vector fields at timelike infinity. Unlike the case for null infinity, the vectors fields at timelike infinity are metric dependent. In order for the vector fields to give faithful representation of the generalized BMS algebra, one needs to use the modified Lie bracket as proposed by Barnich et.al in \cite{barnich}. The algebra is found to be similar to that at null infinity, in which, supertranslation vectors fields form an Abelian subgroup. The (modified) Lie bracket between one supertranslation and a Diff($S^{2}$) vector field is found to be another supertranslation and the algebra between two Diff($S^{2}$) vector fields is found to be another Diff($S^{2}$) vector field.

The natural direction that we would like to pursue after showing the vector field algebra would be to understand the charge algebra at timelike infinity. We expect that a similar modified commutator (just like one uses modified Lie bracket) might be required for understanding the charge algebra. The generalized BMS charge algebra at null infinity was studied in a recent work \cite{miguelalgebra}. It would also be interesting to pursue  along the lines of \cite{anupam, distler} to relate double soft theorems with the generalized BMS charge algebra at timelike infinity. We would like to address these issues in a future work.
\section*{Acknowledgement:} We are thankful to Alok Laddha for posing the problem, for numerous discussions on various subtleties and conceptual issues regarding timelike infinity, as well as for help in preparation of the manuscript. We are thankful to Miguel Campiglia for providing us with crucial inputs at various stages of the project.   We are also thankful to Amitabh Virmani for clearing some of our confusions and informing us about some important references. The work started when Aniket was doing his masters thesis in Chennai Mathematical Institute (CMI) as a part of his BS-MS credit requirement from IISER Pune. He is thankful to CMI for their hospitality. Aniket would also like to thank IISER Pune for giving the permission to stay at CMI in this period and their constant academic support throughout his BS-MS program.
\bigskip
\appendix
%

\section{Variation of Christoffel symbols}\label{delta-gamma}
In this section, we compute the variation of Christoffel symbol under a Diff($S^{2}$) vector field. We start with:
\begin{align}\label{gamma}
	\mathring{\Gamma}^{\alpha}_{\lambda\gamma}=\frac{1}{2}\mathring{h}^{\alpha\eta}\Big(\partial_{\gamma}\mathring{h}_{\lambda\eta}+\partial_{\lambda}\mathring{h}_{\eta\gamma}-\partial_{\eta}\mathring{h}_{\lambda\gamma}\Big)
\end{align}
Now,
\begin{align}\nonumber
	\delta^{g}_{\xi_{SR}}\Big(\mathring{\Gamma}^{\alpha}_{\lambda\gamma}\Big)=\frac{1}{2}\delta^{g}_{\xi_{SR}}&\big(\mathring{h}^{\alpha\eta}\big)\Big(\partial_{\gamma}\mathring{h}_{\lambda\eta}+\partial_{\lambda}\mathring{h}_{\gamma\eta}-\partial_{\eta}\mathring{h}_{\lambda\gamma}\Big)\\
	&+\frac{1}{2}\mathring{h}^{\alpha\eta}\Big(\partial_{\gamma}\delta^{g}_{\xi_{SR}}(\mathring{h}_{\lambda\eta})+\partial_{\lambda}\delta^{g}_{\xi_{SR}}(\mathring{h}_{\eta\gamma})-\partial_{\eta}\delta^{g}_{\xi_{SR}}(\mathring{h}_{\lambda\gamma})\Big)\label{gamma-var}
\end{align}
To evaluate the above expression we
compute the variation of metric by taking the Lie derivative w.r.t. the Diff($S^{2}$) vector
field. Using this, after some algebraic manipulation we finally get \ref{gamma-var} as:
\begin{align}
	\delta^{g}_{\xi_{SR}}\Big(\mathring{\Gamma}^{\alpha}_{\lambda\gamma}\Big)=\frac{1}{2}\big(\mathring{D}_{\gamma}\mathring{D}_{\lambda}+\mathring{D}_{\lambda}\mathring{D}_{\gamma}\big)V_{\mathcal{H}}^{\alpha}+\frac{1}{2}\mathring{h}^{\alpha\eta}~\Big(\mathring{R}_{\lambda\zeta\gamma\eta}+\mathring{R}_{\gamma\zeta\lambda\eta}\Big) V_{\mathcal{H}}^{\zeta}
	\label{gamma-qf}
\end{align}
Here, $\mathring{R}_{\lambda\zeta\gamma\eta}$ and $\mathring{R}_{\gamma\zeta\lambda\eta}$ are the Riemann tensor for the hyperboloid metric $\mathring{h}_{\alpha\beta}$. For the hyperboloid metric we can write the Riemann tensor as:
\begin{align}\label{rh}
 \mathring{R}_{\alpha\beta\gamma\delta}=\mathring{h}_{\alpha\delta}\mathring{h}_{\beta\gamma}-\mathring{h}_{\alpha\gamma}\mathring{h}_{\beta\delta}
 \end{align} 
Substituting \ref{rh} in \ref{gamma-qf} we finally get the variation of Cristofell Symbols as:
\begin{align}
	\delta^{g}_{\xi_{SR}}\Big(\mathring{\Gamma}^{\alpha}_{\lambda\gamma}\Big)=\frac{1}{2}\big(\mathring{D}_{\gamma}\mathring{D}_{\lambda}+\mathring{D}_{\lambda}\mathring{D}_{\gamma}\big)V_{\mathcal{H}}^{\alpha}+\frac{1}{2}V_{\mathcal{H}\gamma}\delta^{\alpha}_{\lambda}+\frac{1}{2}V_{\mathcal{H}\lambda}\delta^{\alpha}_{\gamma}-h_{\gamma\lambda}V^{\alpha}_{\mathcal{H}}\label{gamma-f}
\end{align}
\section{Details of calculation of constraints on the generalized BMS vector fields at timelike infinity}\label{constraints}

In this section, we give the sketch of the calculation that leads to the constraints on the generalized BMS vector  fields \ref{nlc3}, \ref{nlc4}. We start with the gauge condition \ref{nlg}
\begin{align}\label{nlg1}
g^{ab}\partial_{b}\Big(\ln\Big(\sqrt{\frac{h}{\mathring{h}}}~\Big)\Big)+\mathring{\nabla}_{b}g^{ab}=0
\end{align}
\\
If we consider the $\tau$ component of the above expression at leading order in $\tau$ we get:
\begin{align}
\mathring{\nabla}_{\alpha}g^{\tau\alpha}=0
\end{align}
In evaluating the l.h.s of the above expression one can use the non-zero Christoffel symbols for the Minkowski metric
\begin{align}
	\mathring{\Gamma}^{\tau}_{\alpha\beta}=\tau \mathring{h}_{\alpha\beta}~~~~~~~;~~\mathring{\Gamma}^{\alpha}_{\beta\tau}=\tau^{-1}\delta^{\alpha}_{\beta}
\end{align}
to get
\begin{align}\label{ggc1}
(\mathring{h}_{\alpha\beta}h^{(0)\alpha\beta}-3)=0
\end{align}
Similarly, one can find that at the leading order in $\tau$ the  hyperboloid components in \ref{nlg1} evaluates to
\begin{align}\label{ggc2}
h^{(0)\alpha\beta}\partial_{\beta}\big(\ln\big(\sqrt{\frac{h^{(0)}}{\mathring{h}}}\big)\big) + \mathring{\nabla}_{\beta}h^{(0)\alpha\beta}=0
\end{align}
The residual gauge transformations that preserves the above gauge conditions namely \ref{ggc1} and \ref{ggc2} can be found by varying the metric $g_{ab}$ w.r.t. to the vector field as given by \ref{va}.
In both of the expressions we can see that only the metric component $h^{(0)\alpha\beta}$ is involved. One can easily check that $h^{(0)\alpha\beta}$ will be altered only by the $\xi^{(0)\alpha}$ part of the vector field (the $\tau$ component of the vector field only alters the hyperboloid part of the metric at $\mathcal{O}(\tau)$ ). \textit{i.e.} one can see that
\begin{align}
	\mathcal{L}_{\xi}h^{(0)}_{\alpha\beta}=\Big(D_{\alpha}\xi_{\beta}^{(0)}+D_{\beta}\xi_{\alpha}^{(0)}\Big)
\end{align}
Hence, substituting $h^{(0)}_{\alpha\beta}\rightarrow h^{(0)}_{\alpha\beta} + \Big(D_{\alpha}\xi_{\beta}^{(0)}+D_{\beta}\xi_{\alpha}^{(0)}\Big)$ in the gauge conditions \ref{ggc1} and \ref{ggc2} one finally gets the constraints:
\begin{align}\label{nlca3}
2~D^{(\alpha}\xi^{(0)\beta )}\pd_{\beta}\Big(\ln\Big(\sqrt{\frac{h^{(0)}}{\mathring{h}}}~\Big)\Big)~+~ h^{(0) \alpha\beta}\pd_{\beta}D_{\gamma}\xi^{(0)\gamma}~+~2	\mathring{D}_{\beta}D^{(\alpha}\xi^{(0)\beta )}=0\\\label{nlca4}
D^{\alpha}\xi^{(0)\beta} \mathring{h}_{\alpha\beta}=0
\end{align}

%

	\section{Details of calculation for Modified Lie bracket between supertranslation and Diff($S^{2}$) vector field }\label{st-sr-calc}
	In this section, we provide the details of the calculation for the modified bracket between one supertranslation and one Diff($S^{2}$) vector field at $i^{+}$.
	\subsection{Contribution from ordinary Lie bracket}\label{st-sr-ord-calc}
	We start with evaluating the expression \ref{stsrd1}. This can be written as
	\begin{align}
	-(\mathring{\Delta}-&3)\big(V_{\mathcal{H}}^{\alpha} \mathring{D}_{\alpha} f_{\mathcal{H}}\big)\nonumber\\
	&=-\mathring{D}^{\beta}\mathring{D}_{\beta}\big(V_{\mathcal{H}}^{\alpha} \mathring{D}_{\alpha} f_{\mathcal{H}}\big)+3 V_{\mathcal{H}}^{\alpha} \mathring{D}_{\alpha} f_{\mathcal{H}}\nonumber
	\\\nonumber
	&=-\big(\Delta V_{\mathcal{H}}^{\alpha} \mathring{D}_{\alpha} f_{\mathcal{H}}+\mathring{D}_{\beta}V_{\mathcal{H}}^{\alpha} \mathring{D}^{\beta}\mathring{D}_{\alpha} f_{\mathcal{H}}+\mathring{D}^{\beta}V_{\mathcal{H}}^{\alpha}\mathring{D}_{\beta} \mathring{D}_{\alpha} f_{\mathcal{H}}+V_{\mathcal{H}}^{\alpha}\mathring{D}^{\beta}\mathring{D}_{\beta} \mathring{D}_{\alpha} f_{\mathcal{H}}\big)+3V_{\mathcal{H}}^{\alpha} \mathring{D}_{\alpha} f_{\mathcal{H}}\\
	&=-\big(2\mathring{D}^{\beta}V_{\mathcal{H}}^{\alpha}\mathring{D}_{\beta} \mathring{D}_{\alpha} f_{\mathcal{H}}+V_{\mathcal{H}}^{\alpha}\mathring{D}_{\beta}\mathring{D}^{\beta} \mathring{D}_{\alpha} f_{\mathcal{H}}\big)+V_{\mathcal{H}}^{\alpha} \mathring{D}_{\alpha} f_{\mathcal{H}}\label{st-sr-311}
	\end{align}
	We have used $\mathring{\Delta} V_{\mathcal{H}}^{\alpha}=2 V_{\mathcal{H}}^{\alpha}$ in going from third line to the last line. The second term in the above expression \ref{st-sr-311} can be further simplified as
	\begin{align}\label{st2}
	V_{\mathcal{H}}^{\alpha}\mathring{D}_{\beta}\mathring{D}^{\beta} \mathring{D}_{\alpha} f_{\mathcal{H}}&=V_{\mathcal{H}}^{\alpha}\mathring{D}_{\beta}\mathring{D}_{\alpha}\mathring{D}^{\beta}f_{\mathcal{H}}\nonumber
	\\
	&=V_{\mathcal{H}}^{\alpha}\big(\mathring{D}_{\alpha}\mathring{D}_{\beta}\mathring{D}^{\beta}f_{\mathcal{H}}+R^{\beta}_{~\gamma\beta\alpha}\mathring{D}^{\gamma}f_{\mathcal{H}}\big)\nonumber\\
	&=V_{\mathcal{H}}^{\alpha}\mathring{D}_{\alpha} f_{\mathcal{H}}
	\end{align}
	Here, in going from the second line to the third we have used the Riemann tensor $\mathring{R}^{\beta}_{~\gamma\beta\alpha}=\mathring{R}_{\gamma\alpha}=-2\mathring{h}_{\gamma\alpha}$ for $EAdS_{3}$ metric $\mathring{h}_{\alpha\beta}$ and the constraint $\mathring{\Delta} f_{\mathcal{H}}=3f_{\mathcal{H}}$. Using \ref{st2} in \ref{st-sr-311}, we finally get 
	\begin{align}
	-(\Delta-3)\big(V_{\mathcal{H}}^{\alpha} \mathring{D}_{\alpha} f_{\mathcal{H}}\big)=-2\mathring{D}^{\beta}V^{\alpha}_{\mathcal{H}}\mathring{D}_{\beta} \mathring{D}_{\alpha}f_{\mathcal{H}}
	\end{align} 
	\subsection{Contribution from modification terms}\label{st-sr-mod-calc}
	In this section, we evaluate the details of the calculation to arrive at \ref{stsre5}. We have:
	\begin{align}\nonumber
	\delta^{g}_{\xi_{SR}}\Big(\mathring{\Delta}-3\Big)\xi^{\tau}_{ST}&=\delta^{g}_{\xi_{SR}}\Big(\mathring{\Delta}\Big)f_{\mathcal{H}}\\\nonumber
	&=\delta^{g}_{\xi_{SR}}(\mathring{h}^{\alpha\beta}\mathring{D}_{\alpha}\mathring{D}_{\beta})f_{\mathcal{H}}\nonumber\\\nonumber
	&=\delta^{g}_{\xi_{SR}}(\mathring{h}^{\alpha\beta})\mathring{D}_{\alpha}\mathring{D}_{\beta}f_{\mathcal{H}}+\mathring{h}^{\alpha\beta}\delta^{g}_{\xi_{SR}}(\mathring{D}_{\alpha})\mathring{D}_{\beta}f_{\mathcal{H}}+\mathring{h}^{\alpha\beta}\mathring{D}_{\alpha}\delta^{g}_{\xi_{SR}}(\mathring{D}_{\beta})f_{\mathcal{H}}\\
	&=\delta^{g}_{\xi_{SR}}(\mathring{h}^{\alpha\beta})\mathring{D}_{\alpha}\mathring{D}_{\beta}f_{\mathcal{H}}-\mathring{h}^{\alpha\beta}\delta^{g}_{\xi_{SR}}(\Gamma^{\gamma}_{\alpha\beta})\mathring{D}_{\gamma}f_{\mathcal{H}}\label{mstsr1}
	\end{align}
	We have used the fact that variation of the partial derivative term in the covariant derivative does not contribute since this does not depend on the metric. 
	The first term in \ref{mstsr1} can be evaluated by taking the Lie derivative on the hyperboloid metric $\mathring{h}^{\alpha\beta}$ w.r.t $\xi_{SR}$. To evaluate the second term in \ref{mstsr1}, we need the variation of the Christoffel symbols w.r.t the Diff($S^{2}$) vector field. This is computed in Appendix-\ref{delta-gamma} and using this we finally evaluate \ref{mstsr1} as
	\begin{align}\nonumber
	\delta^{g}_{\xi_{SR}}\Big(\mathring{\Delta}-3\Big)&\xi^{\tau}_{ST}\\
	&=-\mathring{D}^{\alpha}V^{\beta}_{\mathcal{H}}(\mathring{D}_{\alpha}\mathring{D}_{\beta}+\mathring{D}_{\beta}\mathring{D}_{\alpha})f_{\mathcal{H}}-\frac{1}{2}\mathring{h}^{\alpha\beta}(\mathring{D}_{\alpha}\mathring{D}_{\beta}+\mathring{D}_{\beta}\mathring{D}_{\alpha})V_{\mathcal{H}\gamma}\mathring{D}_{\gamma}f_{\mathcal{H}}\nonumber\\
	&~~~~~~~~~~~~~~~~~~~~~~~~~~~~~-\frac{1}{2}\mathring{h}^{\alpha\beta}(V_{\mathcal{H}\beta}\delta^{\gamma}_{\alpha}+V_{\mathcal{H}\alpha}\delta^{\gamma}_{\beta})\mathring{D}_{\gamma}f_{\mathcal{H}}+ \mathring{h}^{\alpha\beta}\mathring{h}_{\alpha\beta}V^{\gamma}_{\mathcal{H}}\mathring{D}_{\gamma}f_{\mathcal{H}}\nonumber\\
	&=-\mathring{D}^{\alpha}V^{\beta}_{\mathcal{H}}(\mathring{D}_{\alpha}\mathring{D}_{\beta}+\mathring{D}_{\beta}\mathring{D}_{\alpha})f_{\mathcal{H}}-\frac{1}{2}(\mathring{D}_{\alpha}\mathring{D}^{\alpha}+\mathring{D}_{\alpha}\mathring{D}^{\alpha})V^{\gamma}_{\mathcal{H}}\mathring{D}_{\gamma}f_{\mathcal{H}}\nonumber\\
	&~~~~~~~~~~~~~~~~~~~~~~~~~~~~~~~~~~~~~~-\frac{1}{2}\mathring{D}_{\gamma}f_{\mathcal{H}}\Big(V_{\mathcal{H}\beta}\mathring{h}^{\gamma\beta}+V_{\mathcal{H}\alpha}\mathring{h}^{\gamma\alpha}\Big)+3V^{\gamma}_{\mathcal{H}}\mathring{D}_{\gamma}f_{\mathcal{H}}\nonumber\\
	&=-\mathring{D}^{\alpha}V^{\beta}_{\mathcal{H}}(\mathring{D}_{\alpha}\mathring{D}_{\beta}+\mathring{D}_{\beta}\mathring{D}_{\alpha})f_{\mathcal{H}}-\mathring{\Delta} V^{\gamma}_{\mathcal{H}}\mathring{D}_{\gamma}f_{\mathcal{H}}-\mathring{D}_{\gamma}f_{\mathcal{H}} V^{\gamma}_{\mathcal{H}}+3V^{\gamma}_{\mathcal{H}}\mathring{D}_{\gamma}f_{\mathcal{H}}\nonumber\\
	&
	=-\mathring{D}^{\alpha}V^{\beta}_{\mathcal{H}}(\mathring{D}_{\alpha}\mathring{D}_{\beta}+\mathring{D}_{\beta}\mathring{D}_{\alpha})f_{\mathcal{H}}\nonumber 
	\\
	&=-2\mathring{D}^{\alpha}V^{\beta}_{\mathcal{H}}\mathring{D}_{\alpha}\mathring{D}_{\beta}f_{\mathcal{H}}
	\end{align}
	Hence, we can finally write
	\begin{align}
	\delta^{g}_{\xi_{SR}}\Big(\mathring{\Delta}-3\Big)\xi^{\tau}_{ST}=-2\mathring{D}^{\alpha}V^{\beta}_{\mathcal{H}}\mathring{D}_{\alpha}\mathring{D}_{\beta}f_{\mathcal{H}}
	\end{align}
\section{Details of calculation for Modified Lie bracket of two Diff($S^{2}$) vector fields}\label{srdetails}
\subsection{Contribution from the modification terms}\label{twosrdeltaa2}
In this section, we give the details of the computation of last two terms in \ref{c1}, i.e we evaluate the expression
\begin{align}\label{mref1}
\Big( \mathring{\Delta} -2 \Big)\delta^{g}_{\xi_{SR1}}\xi_{SR2}^{\alpha}-\Big( \mathring{\Delta} -2 \Big)\delta^{g}_{\xi_{SR2}}\xi_{SR1}^{\alpha}
\end{align}
\\
In order to evaluate the above, we start with the variation w.r.t to one of the Diff($S^{2}$) vector field on the gauge condition \ref{nlc4}
\begin{align}\label{a1}
2~D^{(\alpha}\xi^{(0)\beta )}\pd_{\beta}\Big(\ln\Big(\sqrt{\frac{h^{(0)}}{\mathring{h}}}\Big)\Big)~+~ h^{(0) \alpha\beta}\pd_{\beta}D_{\gamma}\xi^{(0)\gamma}~+~2	\mathring{D}_{\beta}D^{(\alpha}\xi^{(0)\beta )}=0
\end{align}
\\
Under variation w.r.t $\xi_{SR1}$ the first term in the above expression becomes 
\begin{align}\nonumber
\delta^{g}_{\xi_{SR1}}\Big(2~D^{(\alpha}\xi^{(0)\beta )}\pd_{\beta}\Big(\ln\Big(\sqrt{\frac{h^{(0)}}{\mathring{h}}}\Big)\Big)\Big)=\delta^{g}_{\xi_{SR1}}\Big(2~D^{ (\alpha}&\xi^{\beta )}\Big)\pd_{\beta}\Big(\ln\Big(\sqrt{\frac{h^{(0)}}{\mathring{h}}}\Big)\Big)\Big)\\
&~~~~~~~~~~~~~+2~D^{ (\alpha}\xi^{\beta )}\pd_{\beta}\Big(D_{\gamma}\xi^{\gamma }_{SR1}\Big)
\end{align}
The r.h.s of the above expression vanishes when one considers the variation of \ref{a1} on the Diff($S^{2}$) vector field $\xi_{SR2}$. This corresponds to evaluating the above expression at $h_{\alpha\beta}^{(0)}=\mathring{h}_{\alpha\beta}$ and $\xi=\xi_{SR2}$.

Consider the variation of the second term in \ref{a1}  w.r.t $\xi_{SR1}$.
\begin{align}\label{sr7}
\delta^{g}_{\xi_{SR1}}\Big(h^{ (0)\alpha\beta}\pd_{\beta}D_{\gamma}\xi^{\gamma}\Big)=\delta^{g}_{\xi_{SR1}}\Big(h^{ (0)\alpha\beta}\Big)\pd_{\beta}D_{\gamma}\xi^{\gamma}+h^{ (0)\alpha\beta}\pd_{\beta}\Big(\delta^{g}_{\xi_{SR1}}\Big(D_{\gamma}\xi^{\gamma}\Big)\Big)
\end{align} 
As we have done previously, the first term in the r.h.s of the above expression will vanish when we finally substitute $h_{\alpha\beta}^{(0)}=\mathring{h}_{\alpha\beta}$, due the divergence free condition of $\xi^{\gamma}_{SR2}$. The second term in \ref{sr7} can be evaluated when $h_{\alpha\beta}^{(0)}=\mathring{h}_{\alpha\beta}$ and $\xi=\xi_{SR2}$ as
\begin{align}\nn
h^{(0)\alpha\beta}\pd_{\beta}\Big(\delta^{g}_{\xi_{SR1}}\Big(D_{\gamma}\xi^{\gamma}_{SR2}\Big)\Big)\Bigr\rvert_{h_{\alpha\beta}^{(0)}=\mathring{h}_{\alpha\beta}}&=\mathring{h}^{ \alpha\beta}\pd_{\beta}\Big(\delta^{g}_{\xi_{SR1}}\big(D_{\gamma}\big)\xi^{\gamma}_{SR2}\Big)\Bigr\rvert_{h_{\alpha\beta}^{(0)}=\mathring{h}_{\alpha\beta}}+\mathring{h}^{\alpha\beta}\pd_{\beta}\Big(D_{\gamma}\delta^{g}_{\xi_{SR1}}\big(\xi^{\gamma}_{SR2}\big)\Big)\Bigr\rvert_{h_{\alpha\beta}^{(0)}=\mathring{h}_{\alpha\beta}}\\\nn
&=\mathring{h}^{ \alpha\beta}\pd_{\beta}\Big(\delta^{g}_{\xi_{SR1}}\big(\mathring{D}_{\gamma}\big)\xi^{\gamma}_{SR2}\Big)+\mathring{h}^{\alpha\beta}\pd_{\beta}\Big(\mathring{D}_{\gamma}\delta^{g}_{\xi_{SR1}}\big(\xi^{\gamma}_{SR2}\big)\Big)\\\nn
&=\mathring{h}^{ \alpha\beta}\pd_{\beta}\Big(\delta^{g}_{\xi_{SR1}}\big(\mathring{\Gamma}^{\gamma}_{\gamma\rho}\big)\xi^{\rho}_{SR2}\Big)+\mathring{h}^{\alpha\beta}\pd_{\beta}\Big(\mathring{D}_{\gamma}\delta^{g}_{\xi_{SR1}}\big(\xi^{\gamma}_{SR2}\big)\Big)\\
&=0+\mathring{h}^{\alpha\beta}\pd_{\beta}\Big(\mathring{D}_{\gamma}\delta^{g}_{\xi_{SR1}}\big(\xi^{\gamma}_{SR2}\big)\Big)\label{step1}\\
&=\mathring{h}^{\alpha\beta}\pd_{\beta}\Big(\Big(\mathring{D}^{\rho}V_{\mathcal{H}}^{\gamma}+\mathring{D}^{\gamma}V_{\mathcal{H}}^{\rho}\Big)\mathring{D}_{\gamma}W_{\mathcal{H}\rho}\Big)\label{step2}
\end{align}
Here, in going from \ref{step1} to \ref{step2} we have used \ref{div1} and \ref{sr5}.

At this point, it will be useful to remember the expression \ref{mref1}. There is a term $\Big( \mathring{\Delta} -2 \Big)\delta^{g}_{\xi_{SR2}}\xi_{SR1}^{\alpha}$ which also needs to be evaluated. This corresponds to doing the same analysis as we have done till now but interchanging $V_{\mathcal{H}}^{\alpha}$ with $W_{\mathcal{H}}^{\alpha}$. This will help us in eliminating many terms which will not appear in the final expression. Therefore, contribution of \ref{step2} corresponding to doing this procedure is equal to 
\begin{align}
h^{(0)\alpha\beta}\pd_{\beta}\Big(\delta^{g}_{\xi_{SR2}}\Big(D_{\gamma}\xi^{\gamma}_{SR1}\Big)\Big)\Bigr\rvert_{h_{\alpha\beta}^{(0)}=\mathring{h}_{\alpha\beta}}=\mathring{h}^{\alpha\beta}\pd_{\beta}\Big(\Big(\mathring{D}^{\rho}W_{\mathcal{H}}^{\gamma}+\mathring{D}^{\gamma}W_{\mathcal{H}}^{\rho}\Big)\mathring{D}_{\gamma}V_{\mathcal{H}\rho}\Big).
\end{align}
Therefore, we get 
\begin{align}
h^{(0)\alpha\beta}\pd_{\beta}\Big(\delta^{g}_{\xi_{SR1}}\Big(D_{\gamma}\xi^{\gamma}_{SR2}\Big)\Big)\Bigr\rvert_{h_{\alpha\beta}^{(0)}=\mathring{h}_{\alpha\beta}}-h^{(0)\alpha\beta}\pd_{\beta}\Big(\delta^{g}_{\xi_{SR2}}\Big(D_{\gamma}\xi^{\gamma}_{SR1}\Big)\Big)\Bigr\rvert_{h_{\alpha\beta}^{(0)}=\mathring{h}_{\alpha\beta}}=0.
\end{align}
Hence, the second term in \ref{a1} will not contribute.\\
We are now left with the variation of the third term in \ref{a1}
\begin{align}\label{sr8}
\delta^{g}_{\xi_{SR1}}\Big(2\mathring{D}_{\beta}D^{ (\alpha}\xi^{\beta )}_{SR2}\Big)=2\mathring{D}_{\beta}\Big(D^{ (\alpha}\delta^{g}_{\xi_{SR1}}\xi^{\beta )}_{SR2}+\delta^{g}_{\xi_{SR1}}\big(D^{ (\alpha}\big)\xi^{\beta )}_{SR2}\Big)
\end{align}
\\
The first term in the above expression evaluated at $h_{\alpha\beta}^{(0)}=\mathring{h}_{\alpha\beta}$ and $\xi=\xi_{SR2}$ can be written as
\begin{align}\label{sr9}
2\Big(\mathring{D}_{\beta}D^{ (\alpha}\delta^{g}_{\xi_{SR1}}\xi^{\beta )}_{SR2}\Big)\Bigr\rvert_{h_{\alpha\beta}^{(0)}=\mathring{h}_{\alpha\beta}}&=\mathring{D}_{\beta}\mathring{D}^{\alpha}\delta^{g}_{\xi_{SR1}}\xi^{\beta}_{SR2}+\mathring{D}_{\beta}\mathring{D}^{\beta}\delta^{g}_{\xi_{SR1}}\xi^{\alpha}_{SR2}\\
&=\mathring{h}^{\alpha\gamma}\mathring{D}_{\gamma}\mathring{D}_{\beta}\delta^{g}_{\xi_{SR1}}\xi^{\beta}_{SR2}+\big(\mathring{\Delta} -2 \big)\delta^{g}_{\xi_{SR1}}\xi^{\alpha}_{SR2}\label{sr10}\\
&=\big(\mathring{\Delta} -2 \big)\delta^{g}_{\xi_{SR1}}W_{\mathcal{H}}^{\alpha}\label{sr11}
\end{align}
\\
The second term in \ref{sr8} evaluated at $h_{\alpha\beta}^{(0)}=\mathring{h}_{\alpha\beta}$ and $\xi=\xi_{SR2}$ can be written as
\begin{align}\label{sr12}
2\mathring{D}_{\beta}\Big(\delta^{g}_{\xi_{SR1}}\big(D^{ (\alpha}\big)\xi^{\beta )}_{SR2}\Big)\Bigr\rvert_{h_{\alpha\beta}^{(0)}=\mathring{h}_{\alpha\beta}}&=\mathring{D}_{\beta}\Big(\delta^{g}_{\xi_{SR1}}\big(D^{\alpha}\big)\xi^{\beta}_{SR2}\Big)\Bigr\rvert_{h_{\alpha\beta}^{(0)}=\mathring{h}_{\alpha\beta}}+\mathring{D}_{\beta}\Big(\delta^{g}_{\xi_{SR1}}\big(D^{\beta}\big)\xi^{\alpha}_{SR2}\Big)\Bigr\rvert_{h_{\alpha\beta}^{(0)}=\mathring{h}_{\alpha\beta}}\\ \nonumber
&=\mathring{D}_{\beta}\Big(\delta^{g}_{\xi_{SR1}}\big(\mathring{D}^{\alpha}\big)\xi^{\beta}_{SR2}\Big)+\mathring{D}_{\beta}\Big(\delta^{g}_{\xi_{SR1}}\big(\mathring{D}^{\beta}\big)\xi^{\alpha}_{SR2}\Big)\\\nonumber
&=\od_{\beta}\Big((\dgv \mathring{h}^{\alpha\gamma})\od_{\gamma}\xi_{SR2}^{\beta}\Big)+\od_{\beta}\Big((\dgv \mathring{h}^{\gamma\beta})\od_{\gamma}\xi_{SR2}^{\alpha}\Big)\\
&~~~~~~~~~+\od_{\beta}\Big(\mathring{h}^{\alpha\gamma}(\dgv \od_{\gamma})\xi_{SR2}^{\beta}\Big)+\od_{\beta}\Big(\mathring{h}^{\beta\gamma}(\dgv \od_{\gamma})\xi_{SR2}^{\alpha}\Big)\label{sr13}
\end{align}
\\
The first two terms in the above expression can be computed using $\dgv \mathring{h}^{\alpha\gamma}=-\Big(\od^{\alpha}V_{\mathcal{H}}^{\gamma}+\mathring{D}^{\gamma}V_{\mathcal{H}}^{\alpha}\Big)$ to get
\begin{align}\nonumber
\od_{\beta}\Big((\dgv \mathring{h}^{\alpha\gamma})\od_{\gamma}&\xi_{SR2}^{\beta}\Big)+\od_{\beta}\Big((\dgv \mathring{h}^{\gamma\beta})\od_{\gamma}\xi_{SR2}^{\alpha}\Big)\\\nonumber
&=-\Big[\od_{\beta}(\od^{\alpha}V_{\mathcal{H}}^{\gamma}+\od^{\gamma}V_{\mathcal{H}}^{\alpha})\od_{\gamma}W_{\mathcal{H}}^{\beta}+\od_{\beta}(\od^{\beta}V_{\mathcal{H}}^{\gamma}+\od^{\gamma}V_{\mathcal{H}}^{\beta})\od_{\gamma}W_{\mathcal{H}}^{\alpha}\\&~~~~~~~~~
+(\od^{\alpha}V_{\mathcal{H}}^{\gamma}+\od^{\gamma}V_{\mathcal{H}}^{\alpha})\od_{\beta}\od_{\gamma}W_{\mathcal{H}}^{\beta}+(\od^{\beta}V_{\mathcal{H}}^{\gamma}+\od^{\gamma}V_{\mathcal{H}}^{\beta})\od_{\beta}\od_{\gamma}W_{\mathcal{H}}^{\alpha}\Big]\label{sr14}
\end{align}

The second term in the above expression can be shown to vanish using \ref{pp1} and \ref{hrt}. The third term can be further simplified using \ref{pp1} and \ref{hrt} to 
\begin{align}
(\od^{\alpha}V_{\mathcal{H}}^{\gamma}+\od^{\gamma}V_{\mathcal{H}}^{\alpha})\od_{\beta}\od_{\gamma}W_{\mathcal{H}}^{\beta}=(\od^{\alpha}V_{\mathcal{H}}^{\gamma}+\od^{\gamma}V_{\mathcal{H}}^{\alpha})\mathring{R}^{\beta}_{\rho\beta\gamma}W_{\mathcal{H}}^{\rho}=-2W_{\mathcal{H}\gamma}(\od^{\alpha}V_{\mathcal{H}}^{\gamma}+\od^{\gamma}V_{\mathcal{H}}^{\alpha})
\end{align}
Therefore, \ref{sr14} can be written as
\begin{align}\nonumber
\od_{\beta}\Big((\dgv \mathring{h}^{\alpha\gamma})\od_{\gamma}&\xi_{SR2}^{\beta}\Big)+\od_{\beta}\Big((\dgv \mathring{h}^{\gamma\beta})\od_{\gamma}\xi_{SR2}^{\alpha}\Big)\\\nonumber
&=-\Big[(\od^{\beta}V_{\mathcal{H}}^{\gamma}+\od^{\gamma}V_{\mathcal{H}}^{\beta})\od_{\beta}\od_{\gamma}W_{\mathcal{H}}^{\alpha}+\od_{\beta}(\od^{\alpha}V_{\mathcal{H}}^{\gamma}+\od^{\gamma}V_{\mathcal{H}}^{\alpha})\od_{\gamma}W_{\mathcal{H}}^{\beta}\\&~~~~~~~~~~~~~~~~~~~~~~~~~~~~~~~~~~~~~~~~~~~~~~~~~~~~-2W_{\mathcal{H}\gamma}(\od^{\alpha}V_{\mathcal{H}}^{\gamma}+\od^{\gamma}V_{\mathcal{H}}^{\alpha})
\Big]\label{sr15}
\end{align}
Now, as we have done previously, the terms in \ref{sr15} that will contribute to \ref{a1} can be found by interchanging $V^{\alpha}$ with $W^{\alpha}$ and ignoring the terms that are same. Finally, the terms that contribute to \ref{a1} in the above expression can be found to be
\begin{align}\nonumber
\Big(\od_{\beta}\Big((\dgv \mathring{h}^{\alpha\gamma})\od_{\gamma}\xi_{SR2}^{\beta}\Big)&+\od_{\beta}\Big((\dgv \mathring{h}^{\gamma\beta})\od_{\gamma}\xi_{SR2}^{\alpha}\Big)\Big)_{\mathrm{non-vanishing}}\\
&=-\od_{\beta}\od^{\alpha}V_{\mathcal{H}}^{\gamma}\od_{\gamma}W_{\mathcal{H}}^{\beta}+2W_{\mathcal{H}\gamma}(\od^{\alpha}V_{\mathcal{H}}^{\gamma}+\od^{\gamma}V_{\mathcal{H}}^{\alpha})-\od_{\beta}\od_{\gamma}W_{\mathcal{H}}^{\alpha}\od^{\beta}V_{\mathcal{H}}^{\gamma}\label{final1}
\end{align}
where, ``non-vanishing" denotes the terms that contribute to \ref{a1}.
Now, let us simplify the last two terms in \ref{sr13}.
\begin{align}\nonumber
\od_{\beta}\Big(\mathring{h}^{\alpha\gamma}(\dgv \od_{\gamma})\xi_{SR2}^{\beta}\Big)&+\od_{\beta}\Big(\mathring{h}^{\beta\gamma}(\dgv \od_{\gamma})\xi_{SR2}^{\alpha}\Big)\\
&=\od_{\beta}\Big(\mathring{h}^{\alpha\gamma}(\dgv \mathring{\Gamma}^{\beta}_{\gamma\rho})\xi_{SR2}^{\rho}\Big)+\od_{\beta}\Big(\mathring{h}^{\beta\gamma}(\dgv \mathring{\Gamma}^{\alpha}_{\gamma\rho})\xi_{SR2}^{\rho}\Big)\label{sr16}
\end{align}
The variation of Christoffel symbol under Diff($S^{2}$) vector field \ref{gamma-f} is
\begin{align}
\delta^{g}_{\xi_{SR1}}\mathring{\Gamma}_{\beta\gamma}^{\alpha}=\frac{1}{2}\Big(\mathring{D}_{\beta}\mathring{D}_{\gamma}+\mathring{D}_{\gamma}\mathring{D}_{\beta}\Big)V_{\mathcal{H}}^{\alpha}+\frac{1}{2}V_{\mathcal{H}\beta}\delta^{\alpha}_{\gamma}+\frac{1}{2}V_{\mathcal{H}\gamma}\delta^{\alpha}_{\beta}-\mathring{h}_{\beta\gamma}V_{\mathcal{H}}^{\alpha}\label{vcs1}.
\end{align}

Let us denote the first term in the above expression involving two covariant derivatives as ``DD" term, the terms containing delta function as ``$\delta$'' term
and the last term as ``h'' term. We can show that, $\delta$ term and $h$ term does not contribute to \ref{sr16}.

The $\delta$ piece contribution of \ref{vcs1} in \ref{sr16} can be evaluated as
\begin{align}
2\od_{\beta}(\mathring{h}^{\alpha\beta}V_{\mathcal{H}\rho}W^{\rho}_{\mathcal{H}})+\od_{\beta}(W^{\alpha}_{\mathcal{H}}V^{\beta}+W^{\mathcal{H}\beta}V^{\alpha}_{\mathcal{H}}),
\end{align} 
which will not contribute because of the similar contribution  when we interchange $V^{\alpha}_{\mathcal{H}}$ with $W^{\alpha}_{\mathcal{H}}$ when evaluating \ref{a1}.\\
The $h$ piece contribution of \ref{vcs1} in \ref{sr16} can be evaluated similarly as 
\begin{align}
\od_{\beta}(V^{\beta}_{\mathcal{H}}W^{\alpha}_{\mathcal{H}}+V^{\alpha}_{\mathcal{H}}W^{\beta}_{\mathcal{H}}),
\end{align}
which will also not contribute when we interchange $V^{\alpha}_{\mathcal{H}}$ with $W^{\alpha}_{\mathcal{H}}$. 
Therefore, we are left with only the contribution of the ``DD" piece which can be written as
\begin{align}\nonumber
\od_{\beta}\Big(\mathring{h}^{\alpha\gamma}&(\dgv \od_{\gamma})\xi_{SR2}^{\beta}\Big)+\od_{\beta}\Big(\mathring{h}^{\beta\gamma}(\dgv \od_{\gamma})\xi_{SR2}^{\alpha}\Big)\\
&=\frac{1}{2}\od_{\beta}\Big(\mathring{h}^{\alpha\gamma}W^{\rho}(\od_{\gamma}\od_{\rho}V^{\beta}_{\mathcal{H}}+\od_{\rho}\od_{\gamma}V^{\beta}_{\mathcal{H}})\Big)+\frac{1}{2}\od_{\beta}\Big(\mathring{h}^{\beta\gamma}W^{\rho}_{
\mathcal{H}}(\od_{\gamma}\od_{\rho}V^{\alpha}_{\mathcal{H}}+\od_{\rho}\od_{\gamma}V^{\alpha}_{\mathcal{H}})\Big)
\end{align} 
Now, using \ref{hrt} the above expression can be written as
\begin{align}\nonumber
\od_{\beta}\Big(\mathring{h}^{\alpha\gamma}&(\dgv \od_{\gamma})\xi_{SR2}^{\beta}\Big)+\od_{\beta}\Big(\mathring{h}^{\beta\gamma}(\dgv \od_{\gamma})\xi_{SR2}^{\alpha}\Big)\\
&=\frac{1}{2}\od_{\beta}\Big(\mathring{h}^{\alpha\gamma}W_{\mathcal{H}}^{\rho}(2\od_{\gamma}\od_{\rho}V_{\mathcal{H}}^{\beta}+\mathring{R}^{\beta}_{\sigma\rho\gamma}V_{\mathcal{H}}^{\sigma})\Big)+\frac{1}{2}\od_{\beta}\Big(\mathring{h}^{\beta\gamma}W_{\mathcal{H}}^{\rho}(2\od_{\gamma}\od_{\rho}V_{\mathcal{H}}^{\alpha}+\mathring{R}^{\alpha}_{\sigma\rho\gamma}V_{\mathcal{H}}^{\sigma})\Big)\nonumber
\\
&=\od_{\beta}\Big(\mathring{h}^{\alpha\gamma}W_{\mathcal{H}}^{\rho}\od_{\gamma}\od_{\rho}V_{\mathcal{H}}^{\beta}\Big)+\od_{\beta}\Big(\mathring{h}^{\beta\gamma}W_{\mathcal{H}}^{\rho}\od_{\gamma}\od_{\rho}V_{\mathcal{H}}^{\alpha}\Big)\nonumber\\&
~~~~~~~~~~~~~~~~~~+\frac{1}{2}\od_{\beta}\Big(2\mathring{h}^{\alpha\beta}W_{\mathcal{H}\rho}V_{\mathcal{H}}^{\rho}\Big)-\frac{1}{2}\od_{\beta}\Big(V_{\mathcal{H}}^{\alpha}W_{\mathcal{H}}^{\beta}+W_{\mathcal{H}}^{\alpha}V_{\mathcal{H}}^{\beta}\Big)\label{sr17}
\end{align}
Only the first term contributes in the above expression when $V^{\alpha}_{\mathcal{H}}$ interchanged with $W^{\alpha}_{\mathcal{H}}$. Therefore, the contribution of \ref{sr17} becomes
\begin{align}\nonumber
&\Big(\od_{\beta}\Big(\mathring{h}^{\alpha\gamma}(\dgv \od_{\gamma})\xi_{SR2}^{\beta}\Big)+\od_{\beta}\Big(\mathring{h}^{\beta\gamma}(\dgv \od_{\gamma})\xi_{SR2}^{\alpha}\Big)\Big)_{\mathrm{non-vanishing}}\\\nonumber
&=\mathring{h}^{\alpha\gamma}\od_{\beta}\Big(W_{\mathcal{H}}^{\rho}\od_{\gamma}\od_{\rho}V_{\mathcal{H}}^{\beta}\Big)+\mathring{h}^{\beta\gamma}\od_{\beta}\Big(W_{\mathcal{H}}^{\rho}\od_{\gamma}\od_{\rho}V_{\mathcal{H}}^{\alpha}\Big)\\
&=\mathring{h}^{\alpha\gamma}\od_{\beta}W_{\mathcal{H}}^{\rho}\od_{\gamma}\od_{\rho}V_{\mathcal{H}}^{\beta}+\mathring{h}^{\beta\gamma}\od_{\beta}W_{\mathcal{H}}^{\rho}\od_{\gamma}\od_{\rho}V_{\mathcal{H}}^{\alpha}+\mathring{h}^{\alpha\gamma}W_{\mathcal{H}}^{\rho}[\od_{\beta},\od_{\gamma}\od_{\rho}]V_{\mathcal{H}}^{\beta}+\mathring{h}^{\beta\gamma}W_{\mathcal{H}}^{\rho}[\od_{\beta},\od_{\gamma}\od_{\rho}]V_{\mathcal{H}}^{\alpha}\nonumber\\&~~~~~~~~~~~~~~~~~~~~~~~~~~~~~~~~~~~~~~~~~~~~~~~~~~~~~~~~~~~~~~~
+\mathring{h}^{\beta\gamma}W_{\mathcal{H}}^{\rho}\od_{\rho}\od_{\beta}\od_{\gamma}V_{\mathcal{H}}^{\alpha}\nonumber\\
&=\od_{\beta}W_{\mathcal{H}}^{\rho}\od^{\alpha}\od_{\rho}V_{\mathcal{H}}^{\beta}+\od^{\gamma}W_{\mathcal{H}}^{\rho}\od_{\gamma}\od_{\rho}V^{\alpha}-5W_{\mathcal{H}}^{\rho}\od^{\alpha}V_{\mathcal{H}\rho}-2W_{\mathcal{H}}^{\rho}\od_{\rho}V_{\mathcal{H}}^{\alpha}\label{final2}
\end{align}
Finally, adding up \ref{final1} and \ref{final2} and interchanging $V_{\mathcal{H}}^{
\alpha}$ with $W_{\mathcal{H}}^{\alpha}$, \ref{a1} evaluates to
\begin{align}\nonumber
\Big(\mathring{\Delta}-2\Big)\dgv  W^{\alpha}_{\mathcal{H}}&-\Big(\mathring{\Delta}-2\Big)\dgw V^{\alpha}_{\mathcal{H}}\\
&=2\Big(\od^{\beta}V^{\gamma}_{\mathcal{H}}\od_{\beta}\od_{\gamma}W^{\alpha}_{\mathcal{H}}-\od^{\beta}W^{\gamma}_{\mathcal{H}}\od_{\beta}\od_{\gamma}V^{\alpha}_{\mathcal{H}}\Big)\nonumber\\
&~~~~~~~~~~~~~~~~~~~~~~~~-2\Big(V_{\mathcal{H}\gamma}\od^{\alpha}W^{\gamma}_{\mathcal{H}}-W_{\mathcal{H}\gamma}\mathring{D}^{\alpha}V^{\gamma}_{\mathcal{H}}\Big).
\end{align}


\begin{thebibliography}{10}
	
\bibitem{miguel-bulk-boundary}
M.~Campiglia, \emph{{Null to time-like infinity Green's functions for
		asymptotic symmetries in Minkowski spacetime}},
\href{https://doi.org/10.1007/JHEP11(2015)160}{\emph{JHEP} {\bfseries 11}
	(2015) 160} [\href{https://arxiv.org/abs/1509.01408}{{\ttfamily
		1509.01408}}].

\bibitem{barnich}
G.~Barnich and C.~Troessaert, \emph{{Aspects of the BMS/CFT correspondence}},
\href{https://doi.org/10.1007/JHEP05(2010)062}{\emph{JHEP} {\bfseries 05}
	(2010) 062} [\href{https://arxiv.org/abs/1001.1541}{{\ttfamily 1001.1541}}].

\bibitem{bondi1}
H.~Bondi, M.~van~der Burg and A.~Metzner, \emph{{Gravitational waves in general
		relativity. 7. Waves from axisymmetric isolated systems}},
\href{https://doi.org/10.1098/rspa.1962.0161}{\emph{Proc. Roy. Soc. Lond. A}
	{\bfseries A269} (1962) 21}.

\bibitem{bondi2}
R.~Sachs, \emph{{Gravitational waves in general relativity. 8. Waves in
		asymptotically flat space-times}},
\href{https://doi.org/10.1098/rspa.1962.0206}{\emph{Proc. Roy. Soc. Lond. A}
	{\bfseries A270} (1962) 103}.

\bibitem{bondi3}
R.~Sachs, \emph{{Asymptotic symmetries in gravitational theory}},
\href{https://doi.org/10.1103/PhysRev.128.2851}{\emph{Phys. Rev.} {\bfseries
		128} (1962) 2851}.

\bibitem{bondi4}
T.~Mädler and J.~Winicour, \emph{{Bondi-Sachs Formalism}},
\href{https://doi.org/10.4249/scholarpedia.33528}{\emph{Scholarpedia}
	{\bfseries 11} (2016) 33528}
[\href{https://arxiv.org/abs/1609.01731}{{\ttfamily 1609.01731}}].

\bibitem{bms-review}
F.~Alessio and G.~Esposito, \emph{{On the structure and applications of the
		Bondi--Metzner--Sachs group}},
\href{https://doi.org/10.1142/S0219887818300027}{\emph{Int. J. Geom. Meth.
		Mod. Phys.} {\bfseries 15} (2018) 1830002}
[\href{https://arxiv.org/abs/1709.05134}{{\ttfamily 1709.05134}}].

\bibitem{ama}
A.~Ashtekar, M.~Campiglia and A.~Laddha, \emph{{Null infinity, the BMS group
		and infrared issues}},
\href{https://doi.org/10.1007/s10714-018-2464-3}{\emph{Gen. Rel. Grav.}
	{\bfseries 50} (2018) 140}
[\href{https://arxiv.org/abs/1808.07093}{{\ttfamily 1808.07093}}].

\bibitem{strom}
T.~He, V.~Lysov, P.~Mitra and A.~Strominger, \emph{{BMS supertranslations and
		Weinberg's soft graviton theorem}},
\href{https://doi.org/10.1007/JHEP05(2015)151}{\emph{JHEP} {\bfseries 05}
	(2015) 151} [\href{https://arxiv.org/abs/1401.7026}{{\ttfamily 1401.7026}}].

\bibitem{strombms}
A.~Strominger, \emph{{On BMS Invariance of Gravitational Scattering}},
\href{https://doi.org/10.1007/JHEP07(2014)152}{\emph{JHEP} {\bfseries 07}
	(2014) 152} [\href{https://arxiv.org/abs/1312.2229}{{\ttfamily 1312.2229}}].

\bibitem{strominger-review}
A.~Strominger, \emph{{Lectures on the Infrared Structure of Gravity and Gauge
		Theory}},  \href{https://arxiv.org/abs/1703.05448}{{\ttfamily 1703.05448}}.

\bibitem{ashokesubleading}
A.~Sen, \emph{{Subleading Soft Graviton Theorem for Loop Amplitudes}},
\href{https://doi.org/10.1007/JHEP11(2017)123}{\emph{JHEP} {\bfseries 11}
	(2017) 123} [\href{https://arxiv.org/abs/1703.00024}{{\ttfamily
		1703.00024}}].

\bibitem{CS}
F.~Cachazo and A.~Strominger, \emph{{Evidence for a New Soft Graviton
		Theorem}},  \href{https://arxiv.org/abs/1404.4091}{{\ttfamily 1404.4091}}.

\bibitem{stromsublead}
D.~Kapec, V.~Lysov, S.~Pasterski and A.~Strominger, \emph{{Semiclassical
		Virasoro symmetry of the quantum gravity $ \mathcal{S}$-matrix}},
\href{https://doi.org/10.1007/JHEP08(2014)058}{\emph{JHEP} {\bfseries 08}
	(2014) 058} [\href{https://arxiv.org/abs/1406.3312}{{\ttfamily 1406.3312}}].

\bibitem{alok:subleading}
M.~Campiglia and A.~Laddha, \emph{{Asymptotic symmetries and subleading soft
		graviton theorem}},
\href{https://doi.org/10.1103/PhysRevD.90.124028}{\emph{Phys. Rev. D}
	{\bfseries 90} (2014) 124028}
[\href{https://arxiv.org/abs/1408.2228}{{\ttfamily 1408.2228}}].

\bibitem{alok-new-symmetries}
M.~Campiglia and A.~Laddha, \emph{{New symmetries for the Gravitational
		S-matrix}}, \href{https://doi.org/10.1007/JHEP04(2015)076}{\emph{JHEP}
	{\bfseries 04} (2015) 076}
[\href{https://arxiv.org/abs/1502.02318}{{\ttfamily 1502.02318}}].

\bibitem{alok-sub-subleading}
M.~Campiglia and A.~Laddha, \emph{{Sub-subleading soft gravitons: New
		symmetries of quantum gravity?}},
\href{https://doi.org/10.1016/j.physletb.2016.11.046}{\emph{Phys. Lett. B}
	{\bfseries 764} (2017) 218}
[\href{https://arxiv.org/abs/1605.09094}{{\ttfamily 1605.09094}}].

\bibitem{alok:gauge-gravity}
M.~Campiglia and A.~Laddha, \emph{{Sub-subleading soft gravitons and large
		diffeomorphisms}}, \href{https://doi.org/10.1007/JHEP01(2017)036}{\emph{JHEP}
	{\bfseries 01} (2017) 036}
[\href{https://arxiv.org/abs/1608.00685}{{\ttfamily 1608.00685}}].

\bibitem{canonical}
G.~Longhi and M.~Materassi, \emph{{A Canonical realization of the BMS
		algebra}}, \href{https://doi.org/10.1063/1.532782}{\emph{J. Math. Phys.}
	{\bfseries 40} (1999) 480}
[\href{https://arxiv.org/abs/hep-th/9803128}{{\ttfamily hep-th/9803128}}].

\bibitem{alok:massive}
M.~Campiglia and A.~Laddha, \emph{{Asymptotic symmetries of gravity and soft
		theorems for massive particles}},
\href{https://doi.org/10.1007/JHEP12(2015)094}{\emph{JHEP} {\bfseries 12}
	(2015) 094} [\href{https://arxiv.org/abs/1509.01406}{{\ttfamily
		1509.01406}}].

\bibitem{barnich2}
G.~Barnich and C.~Troessaert, \emph{{BMS charge algebra}},
\href{https://doi.org/10.1007/JHEP12(2011)105}{\emph{JHEP} {\bfseries 12}
	(2011) 105} [\href{https://arxiv.org/abs/1106.0213}{{\ttfamily 1106.0213}}].

\bibitem{distler}
J.~Distler, R.~Flauger and B.~Horn, \emph{{Double-soft graviton amplitudes and
		the extended BMS charge algebra}},
\href{https://doi.org/10.1007/JHEP08(2019)021}{\emph{JHEP} {\bfseries 08}
	(2019) 021} [\href{https://arxiv.org/abs/1808.09965}{{\ttfamily
		1808.09965}}].

\bibitem{anupam}
A.~Anupam, A.~Kundu and K.~Ray, \emph{{Double soft graviton theorems and
		Bondi-Metzner-Sachs symmetries}},
\href{https://doi.org/10.1103/PhysRevD.97.106019}{\emph{Phys. Rev. D}
	{\bfseries 97} (2018) 106019}
[\href{https://arxiv.org/abs/1803.03023}{{\ttfamily 1803.03023}}].

\bibitem{3d}
C.~Batlle, V.~Campello and J.~Gomis, \emph{{Canonical realization of ( 2+1
		)-dimensional Bondi-Metzner-Sachs symmetry}},
\href{https://doi.org/10.1103/PhysRevD.96.025004}{\emph{Phys. Rev. D}
	{\bfseries 96} (2017) 025004}
[\href{https://arxiv.org/abs/1703.01833}{{\ttfamily 1703.01833}}].

\bibitem{tanabe}
K.~Tanabe and T.~Shiromizu, \emph{{Asymptotic structure at timelike infinity:
		higher orders}},  \href{https://arxiv.org/abs/1103.5183}{{\ttfamily
		1103.5183}}.

\bibitem{ugen}
U.~Gen and T.~Shiromizu, \emph{{Timelike infinity and asymptotic symmetry}},
\href{https://doi.org/10.1063/1.532666}{\emph{J. Math. Phys.} {\bfseries 39}
	(1998) 6573} [\href{https://arxiv.org/abs/gr-qc/9709009}{{\ttfamily
		gr-qc/9709009}}].

\bibitem{miguelalgebra}
M.~Campiglia and J.~Peraza, \emph{{Generalized BMS charge algebra}},
\href{https://arxiv.org/abs/2002.06691}{{\ttfamily 2002.06691}}.
	
\end{thebibliography}
\end{document}